\documentclass[aps,prl,reprint,superscriptaddress]{revtex4-1}
\usepackage[utf8]{inputenc}
\usepackage{natbib}

\usepackage{amsmath}
\usepackage[toc,page]{appendix}
\usepackage{bm}
\usepackage{ amssymb }
\usepackage{graphicx}
\usepackage {epstopdf}
\usepackage{csquotes}
\usepackage{float}
\begin{document}

\title{Impact of bulk-edge coupling on observation of anyonic braiding statistics in quantum Hall interferometers}



\author{J. Nakamura}
\affiliation{Department of Physics and Astronomy, Purdue University}
\affiliation{Birck Nanotechnology Center, Purdue University}


\author{S. Liang}
\affiliation{Department of Physics and Astronomy, Purdue University}
\affiliation{Birck Nanotechnology Center, Purdue University}

\author{G. C. Gardner}
\affiliation{Birck Nanotechnology Center, Purdue University}
\affiliation{Microsoft Quantum Lab Purdue, Purdue University}

\author{M. J. Manfra}
\email[]{mmanfra@purdue.edu}
\affiliation{Department of Physics and Astronomy, Purdue University}
\affiliation{Birck Nanotechnology Center, Purdue University}
\affiliation{Microsoft Quantum Lab Purdue, Purdue University}
\affiliation{School of Electrical and Computer Engineering, Purdue University}
\affiliation{School of Materials Engineering, Purdue University}

\date{\today}

\begin{abstract}
Quantum Hall interferometers have been used to probe fractional charge, and more recently, fractional statistics of quasiparticles. Theoretical predictions have been made regarding the effect of electrostatic coupling on interferometer behavior and observation of anyonic phases. Here we present measurements of a small Fabry-Perot interferometer in which these electrostatic coupling constants can be determined experimentally, facilitating quantitative comparison with theory. At the $\nu = 1/3$ fractional quantum Hall state, this device exhibits Aharonov-Bohm interference near the center of the conductance plateau interrupted by a few discrete phase jumps, and $\Phi_0$ oscillations at higher and lower magnetic fields, consistent with theoretical predictions for detection of anyonic statistics. We estimate the electrostatic parameters $K_I$ and $K_{IL}$ by two methods: by the ratio of oscillation periods in compressible versus incompressible regions, and from finite-bias conductance measurements, and these two methods yield consistent results. We find that the extracted $K_I$ and $K_{IL}$ can account for the deviation of the values of the discrete phase jumps from the theoretically predicted anyonic phase $\theta _a = 2\pi /3$. In the integer quantum Hall regime, we find that the experimental values of $K_I$ and $K_{IL}$ can account for the the observed Aharonov-Bohm and Coulomb dominated behavior of different edge states. 
\end{abstract}
\date{\today}

\maketitle

\section{Introduction}
Fractional quantum Hall states \cite{Tsui1982} are predicted to host exotic quasiparticles exhibiting fractional charge and fractional anyonic braiding statistics \cite{Leinaas1977, Wilczek1982, Laughlin1983, Halperin1984, Arovas1984}. Fractional charge has been demonstrated through shot noise measurements \cite{Heiblum1997, Saminadayar1997, Heiblum2008}, resonant tunneling \cite{Goldman1995Science, Kou2012, McClure2012, Heiblum2017, Ensslin2020-2}, and interferometery \cite{Heiblum2010, Willett2009, Nakamura2019} for numerous fractional quantum Hall states. Recently, experimental evidence for anyonic statistics was demonstrated through quasiparticle collisions \cite{Bartolomei2020} and Fabry-Perot interference \cite{Nakamura2020}. Additionally, evidence of non-Abelian braiding statistics has been seen in interferometry experiments at $\nu=5/2$ \cite{Willett2009, Willett2013, Willett2019}.

Electronic interferometers using quantum point contacts (QPCs) to partition edge states have been proposed as a method to probe both the fractional charge and fractional braiding statistics of quasiparticles \cite{Kivelson1989, Kivelson1990, Gefen1993, Chamon1997}, and substantial theoretical work has been made to understand the behavior of quantum Hall interferometers \cite{Kim2006, Halperin2007, Halperin2011, Rosenow2012, Levkivskyi2012}, including their application to non-Abelian states \cite{Halperin2006, Bonderson2006, Bishara2008, Bishara2009, Stern2010}. For a Fabry-Perot quantum Hall interferometer, the phase difference determining interference will be given by Eqn. \ref{theta_AB} \cite{Chamon1997, Halperin2011}:
\begin{equation} \label{theta_AB}
\frac{\theta}{2\pi}=e^*_{in} \frac{AB}{\Phi_0}+N_L \frac{\theta_a}{2\pi}
\end{equation}

$B$ is the magnetic field, $A$ is the area of the interference path set by the gates, $e^*_{in}$ is the quasiparticle charge on the interfering edge state, $\Phi_0$ is the flux quantum, $N_L$ is the number of localized quasiparticles inside the interferometer, and $\theta_a$ is the anyonic phase associated with the interfering quasiparticles. With QPCs tuned to weak backscattering, oscillations in the conductance across the interferometer will occur with $\delta G \propto \cos (\theta)$, enabling fractional charge and statistics to be probed through transport measurements. For the $\nu = 1/3$ state $\theta_a$ is predicted to be $\frac{2\pi}{3}$ \cite{Arovas1984, Halperin1984, Jain2003, Jain2004, Jain2010}, while different anyonic phases are predicted for different fractional quantum Hall states \cite{JainBook, Halperin2011}. 

An important consideration for operation of Fabry-Perot interferometers is the role of Coulomb interactions. It has been shown that the Coulomb interaction between charge in the bulk of the interferometer and charge at the edge has a major effect on interferometer behavior as it can cause the area of the interferometer to change when charge in the bulk changes, which modifies the Aharonov-Bohm contribution to the interferometer phase \cite{Halperin2011}; this has important effects in both integer and fractional quantum Hall interference. Strong bulk-edge coupling can result in unusual interference behavior, including a decrease in magnetic flux through the interference path when magnetic field is increased, resulting in positively sloped lines of constant phase. For the integer regime this has been referred to as the Coulomb-dominated regime, while behavior where the bulk-edge coupling is weak and interference exhibits the conventional negatively-sloped lines of constant phase has been referred to as the Aharonov-Bohm regime. This definition of Aharonov-Bohm and Coulomb-dominated regimes is not as meaningful for fractional states due to the effect of anyonic statistics \cite{Halperin2021}. Nevertheless, a strong bulk-edge interaction still has critical effects in the fractional regime. Most importantly, strong bulk-edge coupling can make the anyonic phase unobservable \cite{Halperin2011}, making it important to suppress Coulomb charging effects in interferometers \cite{Zhang2009, Heiblum2016-2, Nakamura2019}.

In our previous experiment probing the anyonic phase at $\nu = 1/3$ \cite{Nakamura2020}, the interferometer was in a regime in which the Coulomb charging effects leading to the bulk-edge interaction were highly suppressed, allowing the anyonic phase to be extracted without being greatly obscured. However, this suppression of Coulomb effects also likely results in thermal smearing of quasiparticle transitions in high and low field regions where the bulk becomes compressible.

Here we report on measurements and analysis of interference in a small gate-defined Fabry-Perot interferometer with lithographic dimensions 800 nm $\times$ 800 nm specifically designed to investigate the impact of increased bulk edge coupling. Two quantum point contacts are used to partially backscatter the edge states with interference occurring between the two different backscattered paths. The effective area based on the magnetic field oscillation period at $\nu = 1$ of approximately 21 mT is $\approx 0.2$ $\mu$m$^2$, implying a lateral depletion of approximately 180 nm, similar to previous results for these types of interferometers \cite{Nakamura2019, Nakamura2020}. 

\begin{figure}[h]
\def\ffile{Figs/v=1-3_data_V5.pdf}
\centering
\includegraphics[width=1\linewidth]{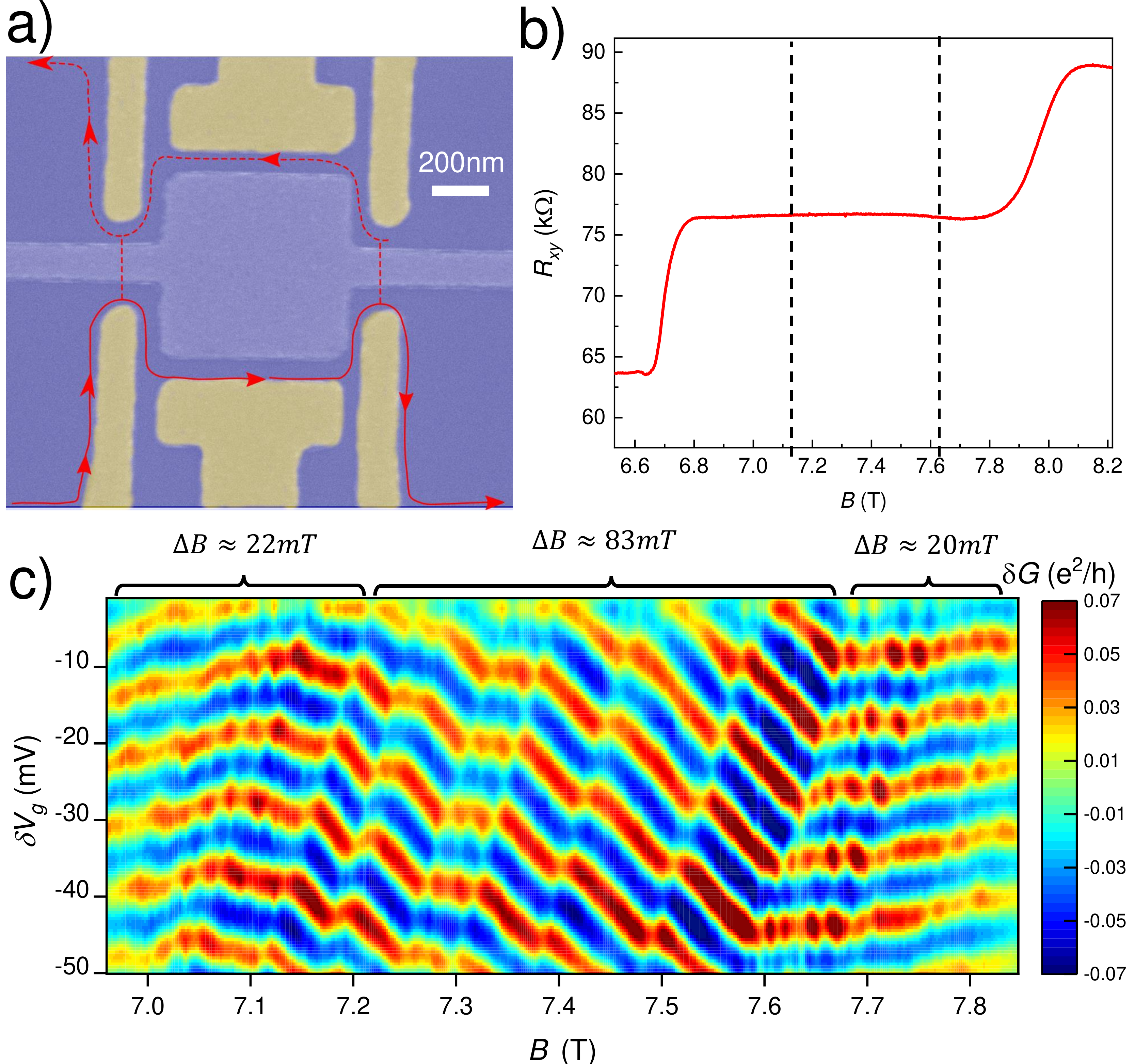}

\caption{\label{v=1-3}Interferometer at the $\nu = 1/3$ fractional quantum Hall state. a) SEM image of interferometer with false color. Yellow regions are the gates which define the interference path. Blue regions corrsepond to the 2DES. There is a center gate in the middle which is kept grounded and does not affect the 2DES density. Red lines indicate edge state trajectories, with backscattered paths denoted by dashed lines. b) Bulk $R_{xy}$ measurement showing the $\nu = 1/3$ conductance plateau. The approximate positions in magnetic field where the interferometer transitions from negatively sloped Aharonov-Bohm behavior to flat lines of constant phase with $\Phi_0$ modulations are marked with dashed lines. c) Interference at $\nu = 1/3$ at a mixing chamber temperature of 10 mK. Near the center there are several discrete jumps in phase which are associated with removal of quasiparticles localized by disorder in the interior of the interferometer. At high and low field the lines of constant phase become nearly independent of magnetic field, but modulations with period approximately $\Phi_0$ can be seen. These modulations are more prominent in the high-field region, particularly close to the transition point at approximately 7.7 T.}
\end{figure}

\section{$\Phi_0$ period modulations in the interference pattern at $\nu = 1/3$}

A key finding from the Rosenow \& Stern model for Fabry-Perot interferometers \cite{Rosenow2020} is that due to the finite energy gap for the creation of quasiparticle and quasihole excitations, there will be a finite range of magnetic field near the center of the state where filling factor $\nu\equiv \frac{n \Phi _0}{B}$ remains fixed and the bulk is incompressible, and in this regime no quasiparticles are created and $\theta$ will evolve primarily due to the Aharonov-Bohm phase. Once the magnetic field has been varied sufficiently far from the center of the state that the chemical potential is outside of the spectral gap, quasiparticles or quasiholes should start to enter the device with the expected period $\Phi_0$. At $\nu = 1/3$ the behavior expected for this regime where the bulk is compressible that is upon the addition of magnetic flux $\Phi_0$, a quasiparticle will be removed from the device (or a quasihole will be added), giving a shift in the anyonic phase of $-2\pi/3$ which cancels out the Aharonov-Bohm contribution to the phase. This results in the leading order interference having no magnetic field dependence, but oscillations still occur as a function of gate voltage \cite{Halperin2011, Halperin2021}. Higher-order contributions to interference are also expected due $\Phi_0$ periodic changes in quasiparticle number \cite{Rosenow2020} but are thermally suppressed. In a previous work at $\nu = 1/3$ \cite{Nakamura2020} the lines of constant phase were observed to shift from a negative slope near the center to zero slope at high and low field, consistent with a shift from constant $\nu$ with an incompressible bulk to constant $n$ with a compressible bulk. $\Phi_0$ modulations from quasiparticle transitions were not observed in the high and low field regions, which was attributed to thermal smearing of the quasiparticle number. Significant thermal smearing is expected due to the small quasiparticle charge $e^* = e/3$ and large screening needed to suppress bulk-edge coupling \cite{Rosenow2020}. 

The device we have measured in this work has an area smaller by approximately a factor of 2 compared to the device in \cite{Nakamura2020}. An SEM image is shown in Fig. \ref{v=1-3}a. Bulk magnetoresistance $R_{xy}$ is shown in Fig. \ref{v=1-3}b indicating the $\nu = 1/3$ state and resistance plateau. Conductance data measured across the interferometer as a function of $B$ and gate voltage variation $\delta V_g$ at $\nu = 1/3$ is shown in Fig. \ref{v=1-3}c ($\delta V_g$ is applied to both side gates, and is relative to -0.8 V). A smooth background is subtracted to emphasize the interference oscillations. For this measurement the QPCs were individually tuned to approximately 90\% transmission at the center of the $\nu = 1/3$ state to achieve the weak backscattering regime. The overall behavior is similar to \cite{Nakamura2020}: near the center of the plateau the lines of constant phase have a negative slope, which is interrupted by several discrete jumps in phase. At low field and high field the lines of constant phase flatten out, consistent with transitions to a compressible bulk with populations of quasiparticles (at low field) or quasiholes (at high field). Unlike \cite{Nakamura2020}, however, there are additional modulations in the interference pattern in the low and high field regions which have a period of $\approx 22$ mT in the low field region and $\approx$ 20 mT in the high field region (Fourier transforms illustrating these periods are shown in Supplemental Fig. 2 and discussed in Supplemental Section 2). This period is close to the Aharonov-Bohm period of $\approx 21$ mT at the integer state $\nu = 1$, indicating that the modulations in the low and high field regions have close to $\Phi_0$ period as predicted in \cite{Rosenow2020}. This suggests that the reduction in device size has enabled quasiparticle transitions and associated higher-order interference terms to be partially resolved in the compressible regions, giving additional experimental evidence of anyonic statistics. Repeatability of the data is discussed in Supplemental Section 1 and Supplemental Fig. 1.


\begin{figure}[h]
\def\ffile{Figs/v=1-3_high_temp.pdf}
\centering
\includegraphics[width=1\linewidth]{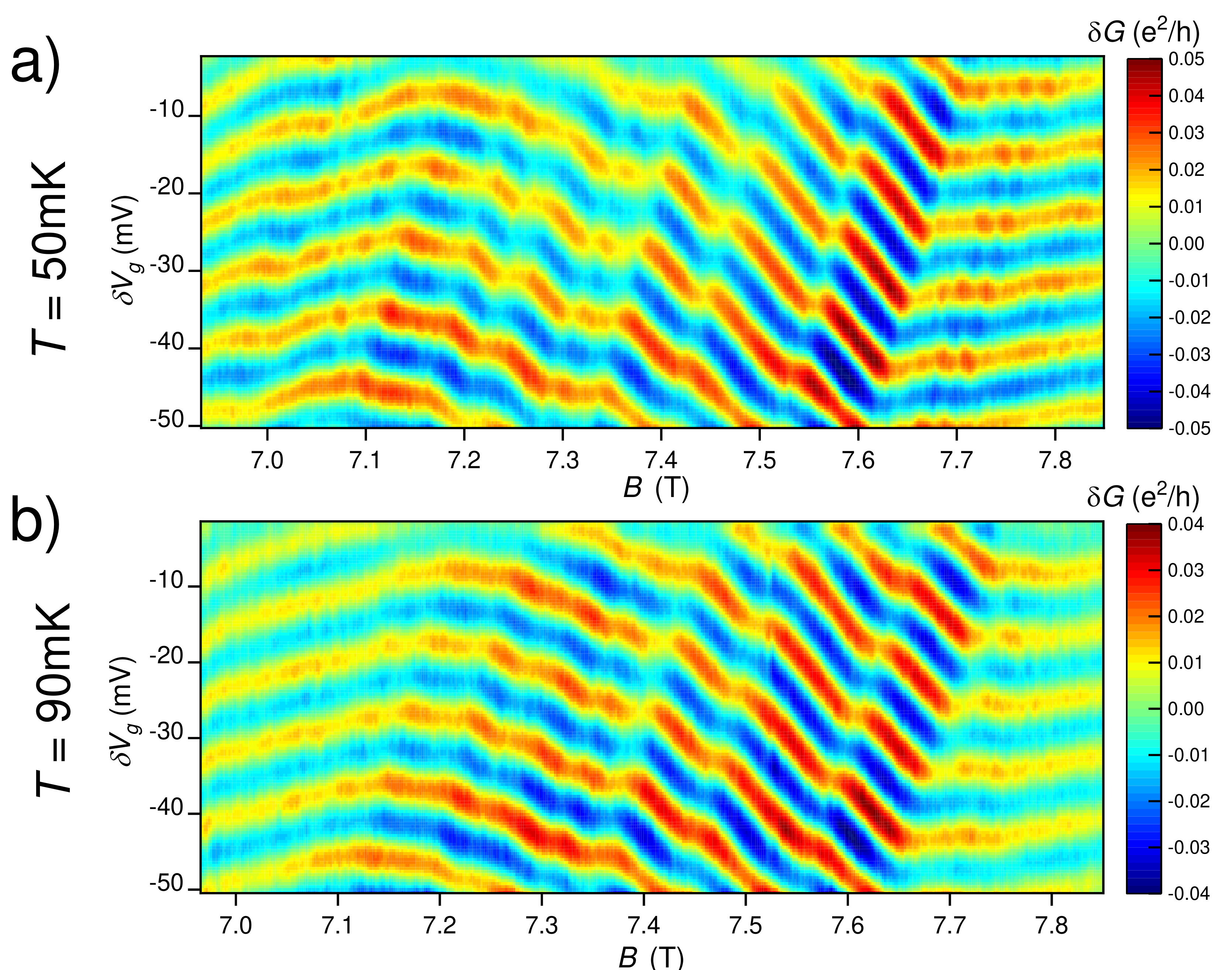}
\caption{\label{v=1-3_hot}Interference at elevated $T$. a) Interference at $\nu = 1/3$ at a mixing chamber temperature of 50 mK. Interference is still of reasonable amplitude, and a few of the discrete jumps in the center are still visible, but the $\Phi_0$ modulations are only weakly visible. b) Interference at 90 mK. The $\Phi_0$ modulations are completely washed out, and the interference amplitude in the low and high field regions are greatly reduced.}
\end{figure}

Interference measurements at elevated temperatures 50 mK and 90 mK are shown in Fig. \ref{v=1-3_hot}a and b. At 50 mK the $\Phi_0$ modulations are greatly suppressed, and at 90 mK they are completely washed out, similar to the observation in \cite{Nakamura2020}. This is consistent with the prediction that the dominant behavior at $\nu = 1/3$ when the bulk is compressible should be oscillations with zero magnetic field frequency \cite{Halperin2011, Halperin2021}, while the $\Phi_0$ modulations are a higher order effect that is more easily thermally smeared. The discrete jumps in the central region also become noticeably less sharp and more smeared out at the higher temperatures.

It is noteworthy in Fig. \ref{v=1-3}c that the $\Phi_0$ modulations are more prominent in the high-field region than in the low-field region, and we have found that this is usually the case in multiple data sets. This may suggest an effect of particle-hole asymmetry in terms of confining quasiparticles inside the interferometer, although we do not have a clear explanation for the effect. Though they carry equal and opposite total charge, quasiparticles and quasiholes have different charge distributions. This can be understood from the composite-fermion picture \cite{Jain1989} from the fact that quasiparticle states involve addition of charge to excited lambda levels \cite{Jain2003A}. How the difference in behavior between quasiparticles and quasiholes might contribute to the more clear $\Phi_0$ modulations in the high field region requires further investigation.

\section{Effect of bulk-edge coupling on interference}
A wide range of interferometer behavior beyond the negatively sloped pure Aharonov-Bohm regime has been observed in previous experiments \cite{Goldman2009, Zhang2009, Heiblum2010, McClure2012, Nakamura2019, Nakamura2020, Ensslin2020}. Theoretical analyses \cite{Halperin2007, Halperin2011, VonKeyserlingk2015, Rosenow2020} have established that electrostatic interaction parameters are crucial in determining the observed behavior, with key parameters being the edge stiffness $K_I$ which describes the energy cost to vary the area of the interfering edge state, and $K_{IL}$ which parameterizes the coupling of the bulk to the edge. Previous experiments have investigated the case of interference when multiple Landau levels are present and inferred the electrostatic parameters governing interference \cite{Heiblum2016-2, Ensslin2020}. Device behavior has been modeled by defining an energy function for the electrostatic energy involving these parameters (Eqn. \ref{Energy}) \cite{Halperin2011}:

\begin{equation} \label{Energy}
E=\frac{K_I}{2}(\delta n_I^2) + \frac{K_L}{2} (\delta n_L)^2+K_{IL}\delta n_L \delta n_I
\end{equation}

In this equation $\delta n_L$ is the variation of the charge in the bulk from the background charge (which includes the quantum Hall condensate density and the contribution from localized charges) and $\delta n_I$ is the variation in charge at the edge from the ideal value. Minimizing the electrostatic energy will result in variations in the area, $\delta A = -\delta n_L \frac{K_{IL}}{K_I} \frac{\Phi_0}{\Delta \nu B}$. Including this variation in area in Eqn. \ref{theta_AB} yields Eqn. \ref{theta_BE} \cite{VonKeyserlingk2015}:

\begin{equation} \label{theta_BE}
\frac{\theta}{2\pi}=e^*_{in} \frac{\Bar{A} B}{\Phi _0}- \frac{K_{IL}}{K_I} \frac{e^*_{in}}{\Delta \nu} (e^*_{in}N_L + \nu_{in} \frac{\Bar{A} B}{\Phi _0}-\Bar{q})+N_L \frac{\theta_a}{2\pi}
\end{equation}

In Eqn. \ref{theta_BE} $\Bar{A}$ is the average area not including the variations $\delta A$ due to the bulk edge coupling, $\Delta \nu$ is the difference between the filling factor corresponding to the interfering edge state and the filling factor of the next-outer edge state which is fully transmitted (for integer states $\Delta \nu = 1$, while for $\nu = 1/3$ $\Delta \nu =1/3$). $\nu_{in}$ is the filling factor corresponding to the interfering edge state, and $\Bar{q}$ is the background charge (which is primarily determined by the ionized donors and may also be changed by the gate voltage). 

\section{Inferring $\frac{K_{IL}}{K_I}$ from magnetic field periods}
Eqn. \ref{theta_BE} implies that in the presence of finite $K_{IL}$, in ranges of magnetic field where the localized quasiparticle number is fixed the the magnetic field oscillation period will be modified from the base value of $\frac{\Phi_0}{e^* \Bar{A}}$. In the presence of bulk-edge coupling this period will change to Eqn. \ref{delta_B} \cite{Halperin2021}:

\begin{equation} \label{delta_B}
\Delta B = \frac{\Phi_0}{e^*_{in} \Bar{A}}(1-\frac{K_{IL}}{K_I}\frac{\nu_{in}}{\Delta \nu})^{-1}
\end{equation}

\begin{figure}[h]
\def\ffile{Figs/v=1_pajama.pdf}
\centering
\includegraphics[width=1\linewidth]{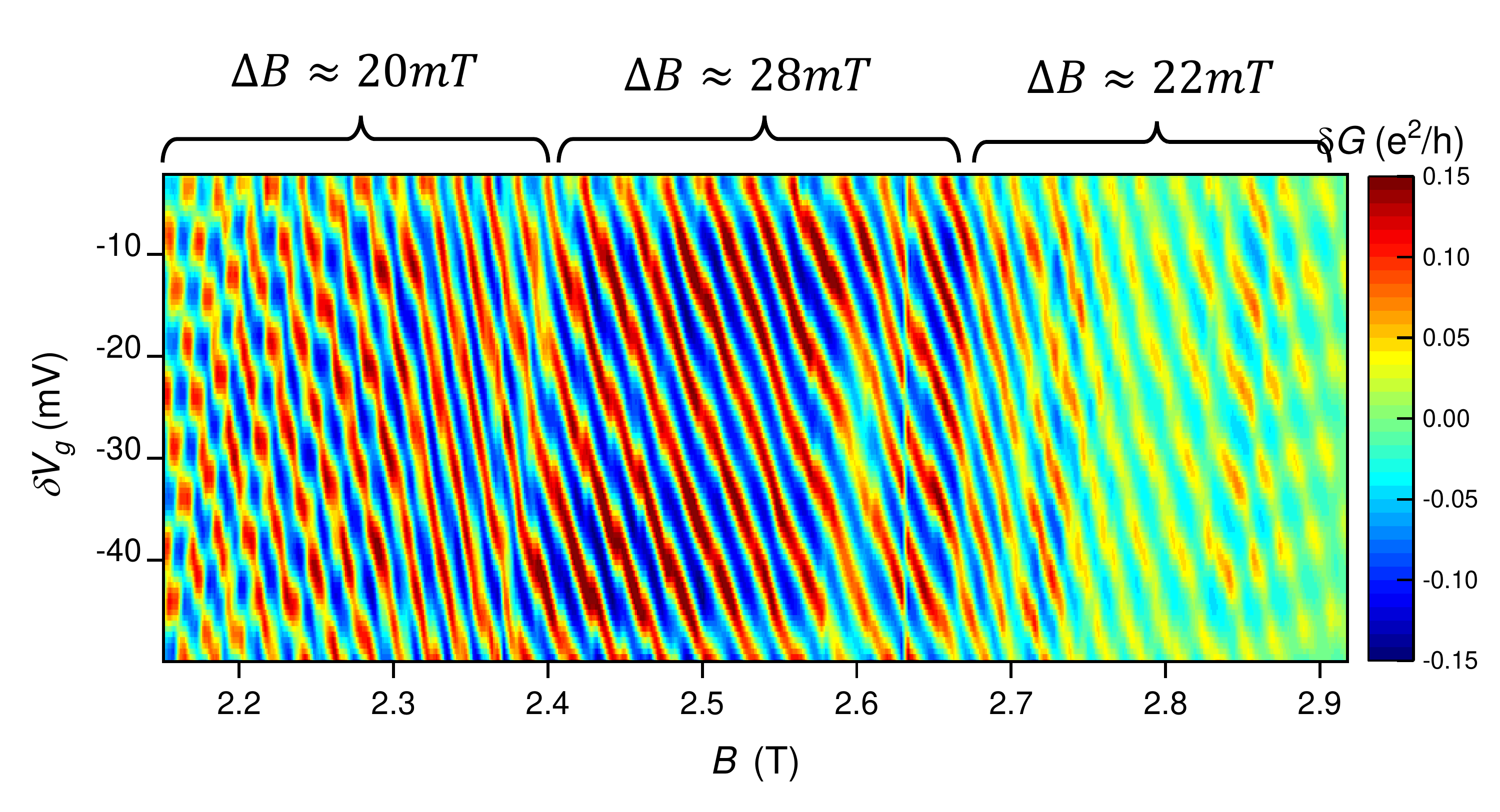}
\caption{\label{v=1_pajama} Interferometer conductance oscillations versus magnetic field and gate voltage at $\nu = 1$. Near the center the magnetic field oscillation period is larger, suggesting an incompressible bulk, while at high and low fields the period becomes smaller (while still maintaining an overall negative slope) suggesting a transition to a compressible bulk.}
\end{figure}

At $\nu = 1$ we observe a region in magnetic field towards the center of the state where the magnetic field period is larger (Fig. \ref{v=1_pajama}), suggesting that in this region $\mu$ is in the gap and the localized electron number is fixed, as predicted by the model in \cite{Rosenow2020}. At higher and lower field the period becomes smaller, suggesting that localized holes and electrons are being added and bringing the period back to the base value of $\frac{\Phi_0}{e^* \Bar{A}}$. This allows $\frac{K_{IL}}{K_I}$ to be extracted based on the ratio of the periods, yielding a value of 0.25 (see Supplemental Section 3 and Supplemental Fig. 3). Similar analysis at $\nu = 1/3$ yields $\frac{K_{IL}}{K_I} = 0.24$. These values imply a moderate effect of bulk-edge coupling at $\nu = 1$ and $\nu = 1/3$.

\section{Extracting $K_I$ and $K_{IL}$ from finite-bias measurements}
Estimating $\frac{K_{IL}}{K_I}$ from Eqn. \ref{delta_B} gives only the ratio of the two parameters, and gives limited insight into the relevant factors which contribute to each term. To extend this analysis and estimate the magnitude of $K_I$ and $K_{IL}$ in our device, we adopt a simple picture in which the energy is a combination of the single-particle energy $E_{sp}$ (which is determined by the external electrostatic confining potential) and the electron interaction energy $E_{int}$ which we estimate by approximating the device as a quantum dot (and extract from zero-field Coulomb blockade measurements). The total energy is a combination of the two terms, $E=E_{sp}+E_{int}$.

In this approximation we estimate the interaction energy by assuming that it can be treated as zero-dimensional quantum dot-like object, so that $E_{int} = \frac{\delta q_{total}^2}{2C}$. $\delta q_{total}$ is the combined bulk and edge excess charge, $\delta q_{total}= e\delta n_{I}+e\delta n_L$, and $C$ is the electrostatic self capacitance of the device. This yields :

\begin{equation} \label{Interaction}
E_{int}=\frac{\delta q_{total}^2}{2C}=\frac{e^2}{C}(\frac{\delta n_I^2}{2}+\frac{\delta n_L^2}{2}+\delta n_L \delta n_I)
\end{equation}

\begin{figure}[h]
\def\ffile{Figs/Integer_data_V2.pdf}
\centering
\includegraphics[width=1\linewidth]{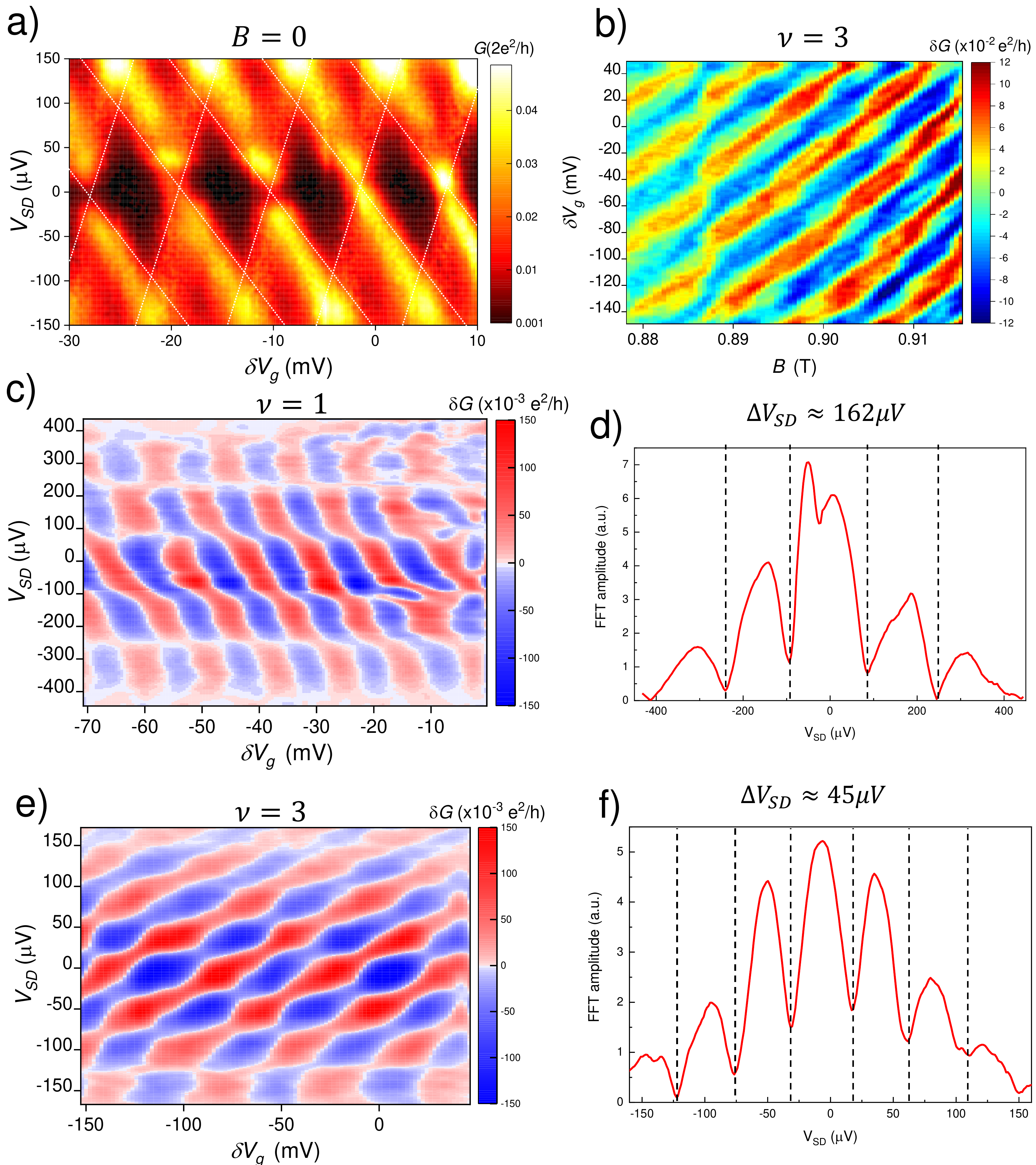}
\caption{\label{IntegerData}Finite-bias measurements. a) Differential conductance measurements at zero magnetic field in the Coulomb blockade regime. The height of the diamond pattern gives the characteristic charging energy $\frac{e^2}{C_{total}}\approx 90$ $ \mu eV$. Note that this energy will include the contribution from the finite DOS, which should be subtracted to yield the electrostatic component that contributes the $K_{IL}$. This yields $\frac{e^2}{2C}\approx 72$ $\mu e$V. b) Interference data for the innermost edge mode at $\nu = 3$. The QPCs are tuned to partially reflect the inner mode and fully transmit the two outer modes. The positive slope indicates that under these conditions the device is in the Coulomb-dominated regime where the bulk-edge interaction is strong. c) Differential conductance measurements at $B = 2.5$ T, the $\nu = 1$ state. The checkerboard pattern suggests that the bias is close to symmetric, although the fact that there is some tilt to the pattern suggests that there is some asymmetry. d) Oscillation amplitude from a FFT versus $V_{SD}$ for $\nu = 1$. The minima in the pattern correspond to the nodes in the checkerboard pattern, allowing extraction of $\Delta V_{SD}\approx 162$ $\mu$V. e) Differential conductance measurement for the inner mode at $\nu  = 3$. f) FFT amplitude versus $V_{SD}$. The spacing between minima gives $\Delta V_{SD} = 45$ $\mu$V, indicating a relatively low velocity which is to be expected for an inner mode.}
\end{figure}

From this we can see that the quantum dot-like charging energy $\frac{e^2}{2C}$ contributes to $K_I$, $K_{IL}$, and $K_L$. $\frac{e^2}{2C}$ can be extracted from the height of Coulomb blockade diamonds. For our device the $B=0$ Coulomb blockade measurements \cite{Beenakker1991} yield $\frac{e^2}{C_{total}}\approx 90 $ $\mu$eV (Fig. \ref{IntegerData}a). This can be refined by subtracting the contribution from the single-particle level spacing due to the finite Density of States (DOS) per unit area $\frac{m^*}{\pi \hbar ^2}$ in 2D at $B$ = 0 which gives a quantum contribution $\frac{e^2}{C_{quantum}}\approx 18$ $\mu e$V for a device with area $\approx 0.2$ $\mu$m$^2$. Then $\frac{e^2}{C}=\frac{e^2}{C_{total}}-\frac{e^2}{C_{quantum}}\approx$ 90 $\mu$eV-$18$ $\mu$eV$=72$ $\mu$eV. 



$E_{sp}$ is set by the confining potential which increases the system energy when area is changed. This external potential can be approximated as a constant electric field $\mathcal{E}$ assuming the variations in area are small. Then, the additional electrostatic potential the charge added to the edge experiences will be $\mathcal{E} \delta l$, where $\mathcal{E}$ is the electric field and $\delta l = \frac{\delta A}{L}$ is the increase in the radius of the interferometer ($L$ is the perimeter of the interference path). $\delta A$ will depend on the amount of charge added to the edge and the sheet density of the interfering edge state, $\delta A = \frac{\delta n_I}{\rho}$ and $\rho = \frac{ \Delta \nu B}{\Phi _0}$. The total change in energy will be the average change in potential times the amount of charge added, $E_{sp}=\frac{e \delta n_I \mathcal{E}\delta l}{2}=\frac{e \delta n_I^2 \mathcal{E} \Phi_0}{2LB \Delta \nu}$. To find the contribution of $E_{sp}$ to $K_I$ we need to determine the value of $\frac{e \mathcal{E} \Phi_0}{LB \Delta \nu}$

The electric field $\mathcal{E}$ also drives the edge velocity, via $\vec{v}_{edge} = \frac{\vec{\mathcal{E}} \times \vec{B}}{B^2}$, so a measure of velocity can be used to extract the electric field and get $E_{sp}$ \cite{Chamon1997}. For integer states (where $e^*=e$ and $\Delta \nu = 1$) and weak backscattering, when the experiment of measuring differential conductance as a function of gate voltage or magnetic field and source drain bias $V_{SD}$ is performed, it exhibits a checkerboard pattern with $\delta G \propto \cos(\frac{2 \pi AB}{\Phi_0}) \cos(\frac{LeV_{SD}}{2 \hbar v_{edge}})$ \cite{McClure2009, Nakamura2019, Deprez2021, Ronen2021} (note that this assumes a symmetric potential drop; the symmetry of the potential drop is discussed in Supplemental Section 4). The product of cosines will result in nodes in the oscillation pattern at $\frac{LeV_{SD}}{2 \hbar v_{edge}}=\pi(n+1/2)$, so that the voltage spacing between nodes $\Delta V_{SD}=\frac{2\pi\hbar v_{edge}}{eL}=\frac{h v_{edge}}{eL}=\frac{\Phi_0 \mathcal{E}}{LB}$. This gives $E_{sp}=\frac{\delta n_I^2}{2} \Delta V_{SD}$.

Differential conductance measurements at $\nu = 1$ are shown in Fig. \ref{IntegerData}c, and the amplitude versus $V_{SD}$ is plotted in Fig. \ref{IntegerData}d. The data exhibits the expected checkerboard pattern, and the spacing between minima in the amplitude (corresponding to the nodes in the oscillation pattern) gives $\Delta V_{SD} \approx 162$ $\mu $V, and $\frac{e \mathcal{E} \Phi_0}{LB}\approx 162$ $\mu e$V. Combining Eqn. \ref{Energy} with the relationships for $E_{sp}$ and $E_{int}$ gives $K_I=\frac{e^2}{C}+\frac{e \mathcal{E} \Phi_0}{LB}\approx 72$ $\mu$eV$+162$ $\mu$eV$=234$ $\mu e$V, while $K_{IL}=\frac{e^2}{C}\approx 72$ $\mu e$V. This gives $\frac{K_{IL}}{K_I}=0.31$. The fact that $\frac{K_{IL}}{K_I}<0.5$ should place the interferometer in the Aharonov-Bohm regime, which is consistent with the observation of predominantly negatively-sloped behavior at $\nu = 1$ (Fig. \ref{v=1_pajama}). Additionally, this value of $\frac{K_{IL}}{K_I}$ is close to the value of 0.25 extracted from Eqn. \ref{delta_B}, giving further validation for the model.

\section{Strong bulk-edge coupling at $\nu=3$}
Strong bulk-edge coupling causes the area of the interference path to decrease when the magnetic field is increased (or when localized quasiparticles are added to the interior of the device), resulting in a positive slope to constant phase lines when $\frac{K_{IL}}{K_I}>0.5$, which has been observed in some previous experiments \cite{Goldman2009, Zhang2009, Heiblum2010, McClure2012}. With $B$ set to $\nu = 3$, interference data is shown for the innermost mode in Fig. \ref{IntegerData}b, where the QPCs are tuned to backscatter the innermost mode and fully transmit the outer modes. There is an overall positive slope to the data, indicating that unlike at $\nu = 1$, the device is in the Coulomb dominated regime and the bulk-edge interaction is strong. This is also supported by the fact that the magnetic field oscillation period is approximately $\frac{\Phi_0}{2}$, similar to previous experiments \cite{Zhang2009, Heiblum2010} and theory \cite{Halperin2007, Halperin2011} for interference of an inner mode at $\nu = 3$.


Differential conductance measurements at $\nu = 3$ are shown in Fig. \ref{IntegerData}e, and the oscillation amplitude versus $V_{SD}$ shown in Fig. \ref{IntegerData}f. From this data $\Delta V_{SD}=45$ $\mu$V, implying a lower velocity and smaller edge stiffness than at $\nu = 1$. This low velocity can be understood from the fact that an inner edge state is being interfered; the inner edge state will be positioned at a region with a more shallow confining potential, resulting in a lower electric field and thus lower velocity \cite{Chklovskii1992, Chklovskii1993, Harshad2018, Nakamura2019}, making it easier for the area enclosed by this edge state to change. The measured value yields $K_I = \frac{e^2}{C}+e\Delta V_{SD}= 72$ $\mu e$V$+45$ $\mu e$V$=117$ $\mu e$V and $K_{IL}=\frac{e^2}{C}\approx 72$ $\mu e$V. This gives $\frac{K_{IL}}{K_I}$=0.62. Since $\frac{K_{IL}}{K_I}>0.5$, the device is predicted to be in the Coulomb-dominated regime, consistent with the observed interference behavior in Fig. \ref{IntegerData}b. This gives additional validation for this method of calculating $K_I$ and $K_{IL}$ since this model is able to correctly predict Coulomb-dominated behavior. On the other hand the outer modes at $\nu = 3$, corresponding to the two spin configurations of the $N = 0$ Landau level, exhibit negatively sloped behavior, indicating that a higher velocity due to a steeper confining potential farther out at the edge gives a stronger $K_I$ (see Supplemental Fig. 5). The outermost edge also exhibits the period-halving effect which has been observed previously \cite{Heiblum2015, Heiblum2018, Ashoori2020} and attributed to inter-edge interaction \cite{Frigeri2019} or electron pairing \cite{Frigeri2020}.  

Recent experiments have extended quantum Hall interferometery to graphene \cite{Deprez2021, Ronen2021}, and provide another opportunity to apply this method for using $K_I$ and $K_{IL}$ to understand Aharonov-Bohm versus Coulomb dominated behavior. In Ref. \cite{Deprez2021} the smallest device with area 3.1 $\mu$m$^2$ had estimated charging energy $E_c = 18$ $\mu e$V and $\Delta V_{SD} \approx 70$ $\mu$V from differential conductance measurements when interfering the outer edge state at $\nu = 2$. This yields $\frac{K_{IL}}{K_I} \approx \frac{18\mu V}{18\mu V + 70 \mu V} \approx 0.2$, placing the device in the Aharonov-Bohm regime, which is consistent with the negatively sloped lines of constant phase observed. Similarly, the 3 $\mu$m$^2$ device in Ref. \cite{Ronen2021} had an estimated $E_c = 16$ $\mu e$V and $\Delta V_{SD} =50$ $\mu$V for the inner mode at $\nu = 2$, yielding $\frac{K_{IL}}{K_I} \approx 0.24$, also in concordance with the observed negatively-sloped Aharonov-Bohm behavior. This suggests that the method for analyzing $K_I$ and $K_{IL}$ can also be applied to graphene devices, which are promising for probing exotic statistics.

\section{Finite bias measurements at $\nu = 1/3$}

The finite-bias behavior of interference at the $\nu = 1/3$ state is expected to be modified by Luttinger liquid effects \cite{Chamon1997, Smits2013}. Ref. \cite{Chamon1997} analyzes the current through interferometers as a function of $V_{SD}$ and $T$, and finds that while integer states should have  uniform spacing of nodes (as discussed in the previous section), for fractional states the innermost nodes will have a narrower spacing than the outer nodes. At low temperature, the positions of the nodes will approximately be given by Eqn. \ref{FractDCNodes}:

\begin{equation} \label{FractDCNodes}
V_{SD} = \frac{h v_{edge}}{e^* L}(n+\frac{1+g}{2}),  n = 0, 1, 2, ...
\end{equation}

Here $g$ is the tunneling exponent, expected to be 1/3 for the $\nu = 1/3$ state. This implies that the innermost nodes will have a narrower spacing than the outer ones, which will have a spacing $\Delta V_{SD}= \frac{h v_{edge}}{e^* L}$. At high temperatures, Ref. \cite{Chamon1997} predicts that the innermost nodes will move outward and reach same spacing as the outer nodes, so that node spacing is uniform as in the integer case. Since this theory calculates the total current it is most convenient to work with the DC current oscillation amplitude $\delta I$ rather than the differential conductance; therefore, we have measured both $\delta G$ and $\delta I$ as a function of $V_{SD}$ at $\nu = 1/3$. 


\begin{figure}[t]
\def\ffile{Figs/v=1_3_velocityV4.pdf}
\centering
\includegraphics[width=1\linewidth]{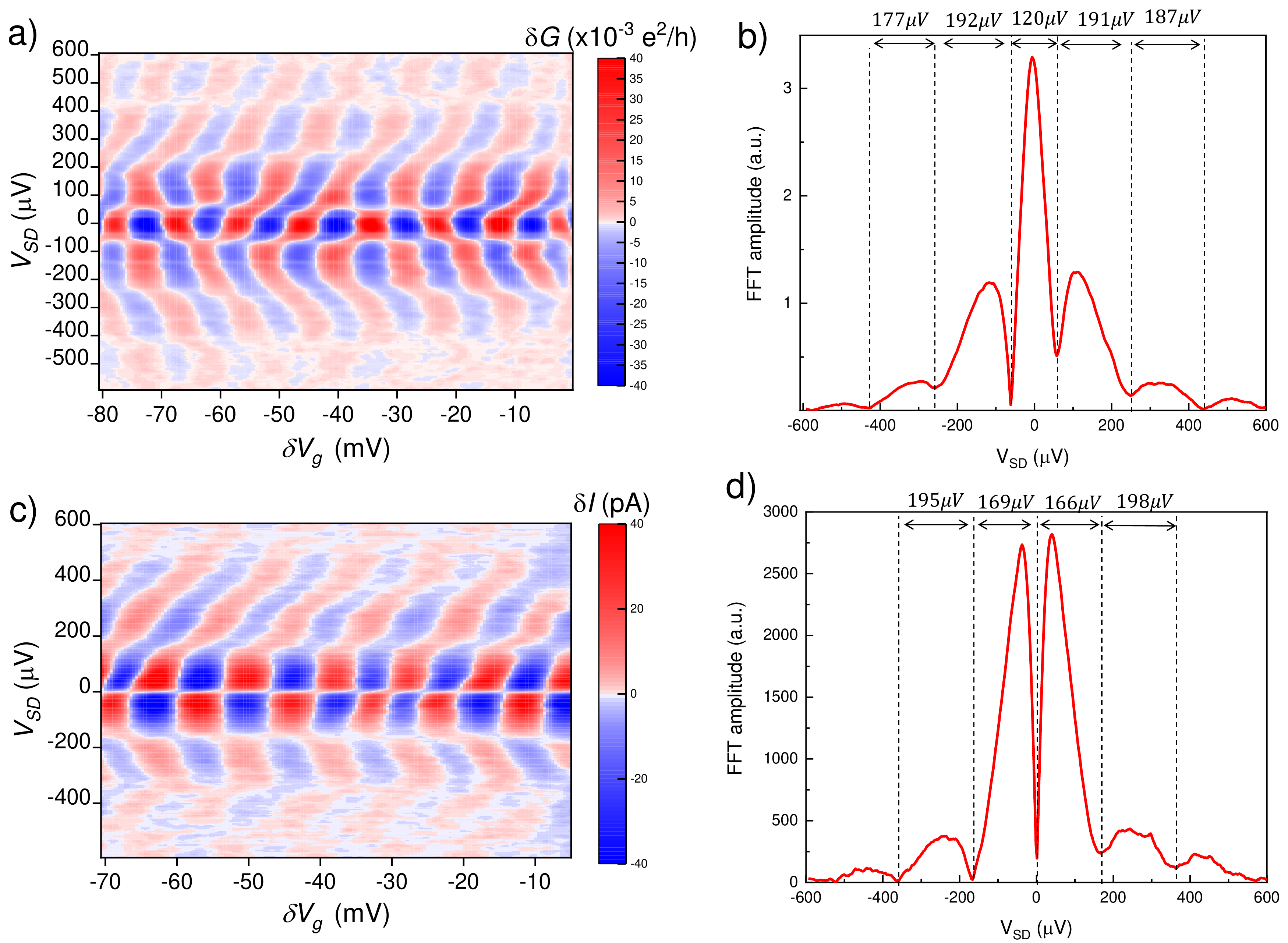}
\caption{\label{v=1-3_velocity_symmetry}Finite-bias measurements at $\nu = 1/3$. a) Differential conductance measurements at $\nu = 1/3$ $B = 7.4$ T. b) Differential conductance oscillation amplitude vs. $V_{SD}$. The spacing between nodes (which appear as minima in the plot) are indicated. It is noteworthy that the spacing between the central nodes is somewhat more narrow than the outer nodes, which may be an indication of Luttinger liquid behavior. c) DC current versus $\delta V_g$ and $V_{SD}$. d) DC current oscillation amplitude versus $V_{SD}$. The minima in the amplitude correspond to nodes in the interference pattern. As predicted by \cite{Chamon1997}, the central nodes have narrower spacing than the outer ones.}
\end{figure}

The differential conductance for $\nu = 1/3$ is plotted in Fig. \ref{v=1-3_velocity_symmetry}a, with the amplitude shown in b. The central separation between the innermost nodes is $\approx 120$ $\mu$V, while the separation between the outer nodes is $\approx 190$ $\mu$V. More direct comparison to \cite{Chamon1997} can be made by to measurements of oscillations in the DC current, shown in Fig. \ref{v=1-3_velocity_symmetry}c and \ref{v=1-3_velocity_symmetry}d. The outer nodes have a separation of $\approx 197$ $\mu $V, while the inner ones have a separation of $\approx 167$ $\mu$V. This is consistent with the expectation that the inner nodes should have a narrower spacing; however, the ratio of the inner to the outer node spacing is 0.85, which is somewhat smaller than the value of 2/3 predicted by Eqn. \ref{FractDCNodes} from \cite{Chamon1997}. A possible explanation for this discrepancy is that the large biases applied in these measurements cause significant heating of the electrons in the device, shifting behavior towards the high-temperature limit of uniform node spacing. Additionally, at elevated mixing chamber temperature the innermost node moves to higher $V_{SD}$, approaching the spacing of the outer nodes as anticipated by \cite{Chamon1997}; see Supplemental Fig. 6 and Supplemental Section 5. The qualitative agreement suggests that this theory can be used to extract $v_{edge}$ from the outer node spacing of $\Delta V_{SD} \approx 197$ $\mu $V (the differential conductance measurement shows similar values for the outer node spacing). Then, $E_{sp} = \frac{e \mathcal{E}\Phi_0}{LB \Delta \nu}=e \Delta V_{SD} \frac{e^*}{e \Delta \nu}$; with $e^* = e/3$ and $\Delta \nu = 1/3$ for the $\nu = 1/3$ state, $E_{sp} = e \Delta V_{SD}$ as for the integer case. Using $\Delta V_{SD} \approx 197$ $\mu$V gives  $K_I\approx 72$ $\mu e$V$+197$ $\mu e$V$=269$ $\mu e$V, while as before $K_{IL}=\frac{e^2}{C}\approx 72$ $\mu e$V. This gives $\frac{K_{IL}}{K_I} = 0.27$, close to the value of 0.24 extracted from Eqn. \ref{delta_B}. A possible limitation to this approach is that the equations for current in Ref. \cite{Chamon1997} were developed without including bulk-edge coupling. Future theoretical work could refine this analysis by analysing bulk-edge coupling corrections to the positions of nodes and improve the accuracy of $K_I$. 

The narrower spacing of the innermost nodes, being a signature of Luttinger-liquid behavior, contrasts with the nearly uniform node spacing observed for the integer states at $\nu = 1$ and $\nu = 3$ in Fig. \ref{IntegerData}a and c, which is expected for Fermi liquids. Previous experimental evidence for Luttinger-liquid behavior of fractional quantum Hall edge states has been seen in tunneling experiments \cite{Chang1996, Chang2003}, while here we have shown evidence through interferometry. 



\section{Phase jumps at $\nu=1/3$}
\begin{figure}[t]
\def\ffile{Figs/v=1_3_phase_analysis_with_dataV3.pdf}
\centering
\includegraphics[width=1.0\linewidth]{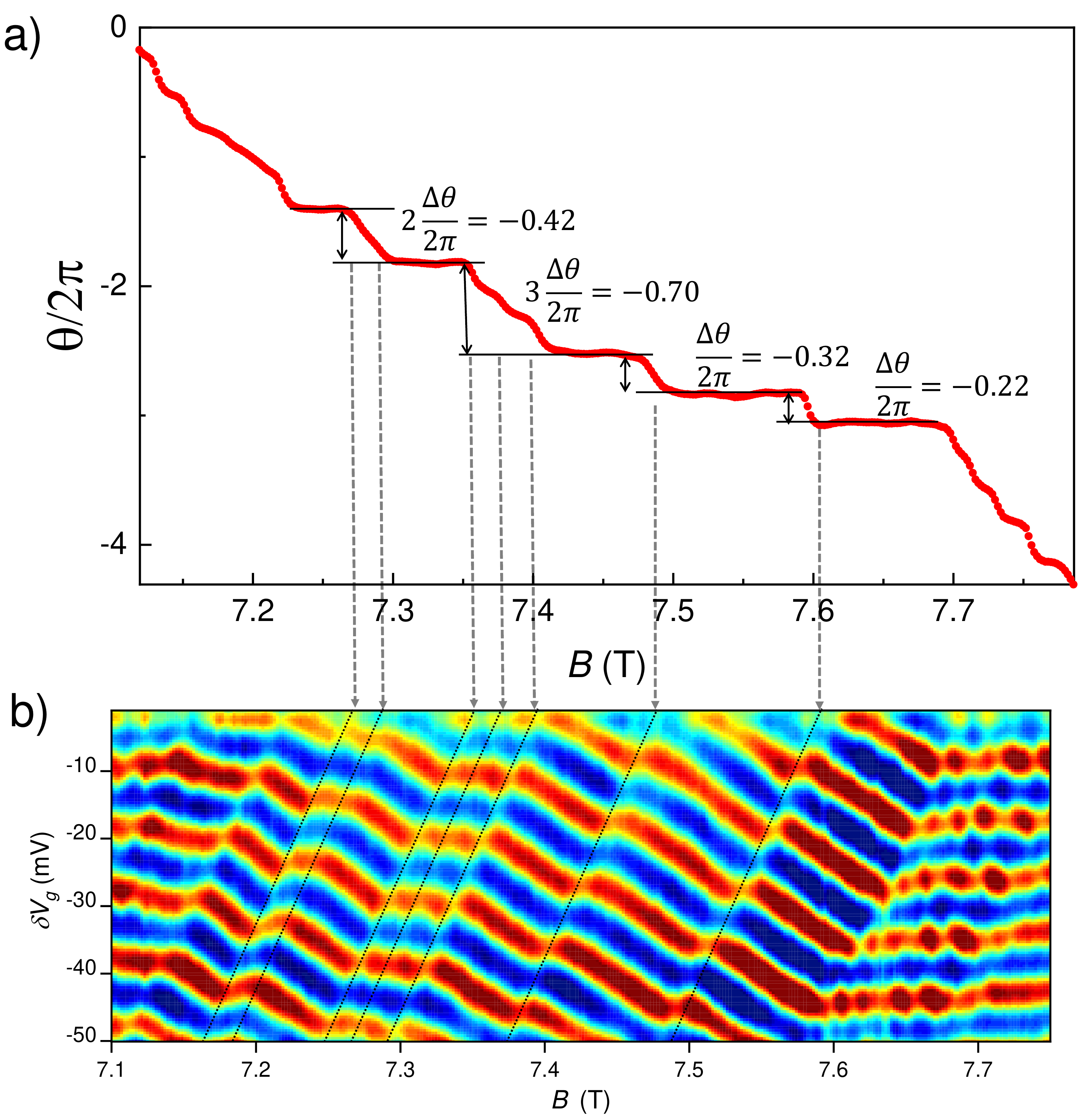}
\caption{\label{phases}Phase versus magnetic field. a) Phases extracted from Fourier transforms of the data in Fig. \ref{v=1-3}a. The FFTs are performed along cuts of the conductance parallel to discrete phase jumps. The phase is evaluated at the peak frequency which corresponds to the Aharonov-Bohm oscillation frequency. The Aharonov-Bohm effect gives a a constant linear change of phase with $B$ which has been subtracted off to yield the contribution from localized quasiparticles, and plateaus in phase occur which correspond to the regions between phase jumps. The change in phase for the discrete jumps are indicated; the leftmost discrete jump appears to correspond to two closeby jumps, while the second from the left appears to consist of three closeby jumps; in these cases the individual phase jumps are not readily resolved, but the total phase change can be calculated and divided into average individual phase changes. b) Raw data (repeated from Fig. \ref{v=1-3}a) indicating where the discrete jumps in a) occur in the data.}
\end{figure}

Several discrete phase jumps can be seen in Fig. \ref{v=1-3}a, similar to previous observations \cite{Nakamura2020}, which may be caused by the anyonic phase when the number of localized quasiparticles inside the interferometer changes. To extract the values of these phase jumps, we have calculated the phase at each value of the magnetic field by taking Fourier transforms along cuts parallel to the lines corresponding to discrete jumps. Then we subtract off the Aharonov-Bohm contribution to the phase (which simply results in continuous phase evolution and a constant linear slope in phase vs. $B$). The process for extracting the phases in this way is discussed in detail in the Supplemental Section 5 and illustrated in Supplemental Fig. 7; this method should enable a more accurate phase extraction than the fitting method in \cite{Nakamura2020} and has the additional advantage of not needing the position of each jump to be specified. The resulting phase after subtracting the Aharonov-Bohm contribution should be due to the anyonic contribtuion, and is plotted in Fig. \ref{phases}a. As can be seen in Fig. \ref{v=1-3}a, some of the phase jumps are very close to each other, so that the individual jumps in phase are not readily resolvable; in particular there appear to be two very close jumps at $\approx 7.28$ T and three close jumps at $\approx 7.37$ T. While the individual phase jumps cannot be isolated, the combined phase jump can be extracted from the data. At low fields (below approximately 7.2 T) and high fields (above approximately 7.7 T) the phase exhibits a staircase pattern due to $\approx \Phi_0$ periodic additions of quasiparticles, although since there is still significant smearing this staircase pattern is not sharp.

The values of the phase jumps (both the individual ones from isolated jumps and the combined ones when multiple are very close) are listed in Fig. \ref{phases}a, and the corresponding part of the data where the jumps occur is indicated in b. These values are calculated by taking the average value of the phase on each plateau and subtracting the adjacent values to get the jump in phase. Averaging all the jumps (and taking into account the fact that some of the changes in phase are most likely due to multiple discrete jumps) yields an average change in phase $\frac{\overline{\Delta \theta}}{2\pi}=-0.24 \pm 0.04$ (uncertainty is estimated from the standard deviation of the phase jumps). This change is smaller than the value of $\theta _a = \frac{2\pi}{3}$ expected from theory (note however that the difference in sign is accounted for the fact that quasiparticles are expected to be removed as field is increased). However, theoretical works \cite{Halperin2011, VonKeyserlingk2015, Halperin2021} predict a modification to the value of the phase jump that occurs when a quasiparticle is added due to bulk edge coupling (Eqn. \ref{theta_correction}): 

\begin{equation} \label{theta_correction}
\frac{\Delta \theta}{2\pi}=\frac{\theta _a}{2\pi}-\frac{K_{IL}}{K_I}\frac{e^{*2}}{\Delta \nu}
\end{equation}

This modification comes about because when a quasiparticle enters the bulk, it's electric charge will cause the area of the interferometer to change, leading to a change in the Aharonov-Bohm phase in addition to the anyonic phase. This correction can be included into the extraction of the anyonic phase at $\nu = 1/3$ by $\frac{\theta _a}{2\pi} = -\frac{\overline{\Delta \theta}}{2\pi}-\frac{1}{3} \frac{K_{IL}}{K_I}$. Using the value of $\frac{K_{IL}}{K_I} = 0.24$ extracted from Eqn. \ref{delta_B} gives $\frac{\theta _a}{2\pi} = 0.32 \pm 0.05$, while using $\frac{K_{IL}}{K_I} = 0.27$ from finite bias measurements yields $\frac{\theta _a}{2\pi} = 0.33 \pm 0.05$; these values are close to the value of $\theta _a = \frac{2\pi}{ 3}$ from theoretical and numerical studies \cite{Arovas1984, Jain2003, Jain2004, Jain2010}. Thus, although the bulk-edge interaction partially obscures the anyonic phase, accounting for this effect indicates that the anyonic phase is close to the theoretically predicted value, giving strong support to the theoretical works \cite{Halperin2011, VonKeyserlingk2015, Halperin2021} and consistent with previous experiments at $\nu = 1/3$ \cite{Bartolomei2020, Nakamura2020}.


\section{Can the phase jumps at $\nu=1/3$ be explained by bulk-edge coupling alone?}
Since bulk-edge coupling can cause discrete jumps in phase even for electrons \cite{Halperin2011, Heiblum2016-2, Ensslin2020}, it should be considered whether the discrete jumps in phase observed at $\nu = 1/3$ can be explained by bulk-edge coupling alone rather than anyonic statistics \cite{Halperin2021}. 

From Eqn. \ref{theta_BE} and Ref. \cite{VonKeyserlingk2015} the change in phase for adding a quasiparticle is $-\frac{K_{IL}}{K_I}\frac{e^{*2} _{in}}{\Delta \nu}$ if $\theta _a$ is assumed to be zero. For $\nu = 1/3$ where $\Delta \nu = 1/3$ and $e^* _{in} = 1/3$, this would result in phase jumps of $\frac{\Delta \theta}{2\pi}=\frac{1}{3} \frac{K_{IL}}{K_I}$ (note the change in sign due to the fact that quasiparticles are being removed by increasing the magnetic field rather than added). Using the value of $\frac{K_{IL}}{K_I}$ of 0.27 extracted from the period measurements (with the value from differential conductance measurements being similar, assuming that the interpretation at $\nu = 1/3$ is correct) gives an expected $\frac{\Delta \theta}{2\pi}=0.09$. This value is significantly different from the phase jumps observed in the data in Fig. \ref{phases}, and is of opposite sign. This suggests that while bulk-edge coupling does reduce the value of the phase jumps observed in this device, the phase jumps cannot be explained by bulk edge coupling alone without anyonic statistics. An assumption made in this analysis is that the charge of the localized quasiparticles is equal to the theoretically predicted value $e^*=e/3$, whereas a larger localized charge would result in a greater phase jump contribution from bulk-edge coupling. Scanning probe experiments have observed $e/3$ localized charge at the $\nu = 1/3$ state \cite{Yacoby2004}, supporting the assumption of fractional charge. Additionally, in previous measurements of a larger device with weak bulk edge coupling discrete phase jumps close to the expected anyonic phase of $2\pi /3$ were observed \cite{Nakamura2020}, slightly larger than the jumps measured in the present device. If the phase jumps were caused only by bulk edge coupling they would be expected to be significantly larger in the smaller device with greater $K_{IL}$. The fact that they are instead slightly smaller is consistent with the anyonic phase being partially obscured by bulk-edge coupling, but not with being caused exclusively by it.    

\section{Conclusions}
In conclusion, we have demonstrated experimental evidence for multiple theoretical predictions of quantum Hall interferometers. We have observed $\Phi_0$ period modulations in interference at the $\nu = 1/3$ state, which are a signature of anyonic statistics when the bulk is compressible. We have demonstrated two approaches for estimating the impact of bulk edge coupling: using the ratio of the magnetic field periods, and extracting the electrostatic coupling constants $K_I$ and $K_{IL}$ directly from finite bias measurements. Uneven node spacing  observed at $\nu = 1/3$ in finite-bias measurements indicates Luttinger liquid behavior. Although our model makes several simplifications, we find that this approach validates theoretical predictions for distinguishing between the Aharonov-Bohm and Coulomb dominated regimes in the integer quantum Hall regime. Accounting for the correction to $\Delta \theta$ from finite $\frac{K_{IL}}{K_I}$ yields values  of $\theta_ {a}$ in agreement with the theoretically predicted value at $\nu = 1/3$, supporting previous experiments. An important finding is that the parameter $\frac{K_{IL}}{K_I}$ can vary between different edge states in the same device, which makes inner edge states more likely to be Coulomb dominated. This work will inform future experimental and theoretical analysis of quantum Hall interferometry. 

\section{Methods}
This interferometer utilizes a high mobility GaAs/AlGaAs heterostructure grown by molecular beam epitaxy \cite{Manfra2014, Gardner2016}. The bulk electron density is approximately $0.6 \times 10^{11}$cm$^{-2}$ and mobility is $3.2 \times 10^6$ cm$^2$V$^{-1}s^{-1}$. The structure also includes additional screening wells with a setback of 25 nm from the main quantum well to reduce the charging energy and bulk-edge coupling so that anyonic statistics can be observed. The screening well design may also enhance the steepness of the confining potential \cite{Harshad2018, Nakamura2019}, which may be important for preventing edge reconstruction that may lead to dephasing by neutral modes \cite{Heiblum2019, Hu2009, Gefen2016}. The screening well heterostructure design has been described in detail in Ref. \cite{Nakamura2019}. Though the structure has the same layer stack as the one  in Ref. \cite{Nakamura2020}, the wafer is different and was grown at a different time. 

Optical lithography and wet etching to define the mesa. Ni/Au/Ge Ohmic contacts were deposited and annealed to make electrical contact to the 2DEG. Electron beam lithography and electron beam evaporation (5nm Ti/10nm Au) were used to define the interferometer gates. Optical lithography and electron beam evaporation (20nm Ti/150nm Au) were used to define bondpads and the surface gates around the Ohmic contacts. The substrate was mechanically polished to to make it thin enough to define metal backgates to deplete electrons in the bottom screening well, which were patterned by optical lithography and deposited by electron beam evaporation (100nm Ti/150nm Au).

Measurements are performed using standard voltage-biased low frequency lock-in amplifier techniques with a typical excitation of 5 $\mu$V and frequency of 37 Hz in a dilution refrigerator with a base mixing chamber temperature of 10 mK.

\section{Acknowledgements}
 This work is supported by the U.S. Department of Energy, Office of Science, Office of Basic Energy Sciences, under award number DE-SC0020138. G. C. Gardner and M. J. Manfra acknowledge support from Microsoft Quantum.  The authors thank Bertrand Halperin for valuable suggestions to improve the manuscript. M. J. Manfra thanks Dima Feldman and Steve Simon for fruitful discussions during the early phases of this project.

\end{document}


\renewcommand{\figurename}{SUPP. FIG.}

\title{Supplementary Material for ``Impact of bulk-edge coupling on observation of anyonic braiding statistics in quantum Hall interferometers''}
\date{\today}

\maketitle

\section{Supplemental Section 1: Repeatability of Data at $\nu = 1/3$}

\begin{figure}[h]
\def\ffile{Figs/v=1-3_repeatability.pdf}
\centering
\includegraphics[width=1\linewidth]{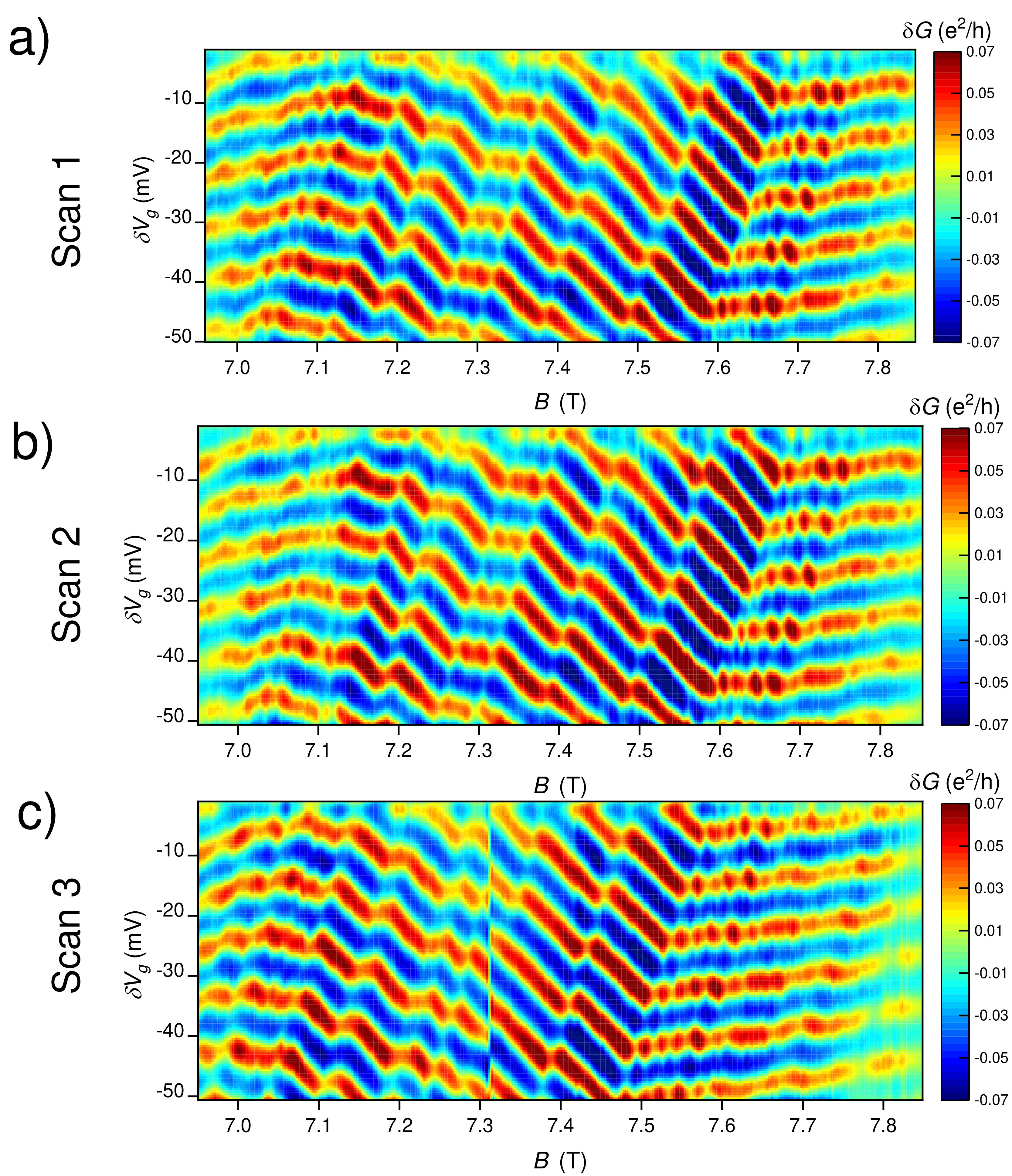}
\caption{\label{repeat} a) 1st scan of conductance versus magnetic field and gate voltage at $\nu = 1/3$. This is the same data presented in the main text. b) 2nd scan over the same magnetic field range using the same paramaters. c) 3rd scan across the same region. In this third scan a clear switching event occurs at approximately 7.3 T. }
\end{figure}

The behavior observed at $\nu = 1/3$ was found to be repeatable upon subsequent measurements. This is shown in Fig. \ref{repeat}, with panels a, b, and c showing conductance versus gate voltage in three separate scans taken one after the other (the first scan is shown in the main text). The discrete jumps in the central region and behavior in the high and low field region are repeatable in each scan. Each scan takes approximately 3 hours. Notably, in the third scan a switching event is visible at approximately 7.3 T. 

Measurements are taken by sweeping gate voltage from the starting value to the final value (from higher to lower voltage) while measuring the conductance, returning the gate voltage to the starting value, then taking a step in magnetic field of 2 mT and repeating the process across the whole magnetic field range. Therefore, changes in the electrostatic potential due to charge noise (which is common in dopeed GaAs/AlGaAs heterostructures) should result in vertical discontinuities in the conductance pattern, such as the one visible in Scan 3. Though we observe occasional jumps like this in our device, they are relatively rare, with only one such jump visible in the approximately 3 hour measurement of Scan 3, and none visible in Scans 1 or 2, which indicates fairly stable device operation. We have used the technique of bias cooling in which a positive bias of +600 mV is applied to the gate when cooling from room temperature, which enables smaller gate voltages to be used and thus reduces charge noise. It is noteworthy that the discrete jumps in phase highlighted in the main text attributed to changes in localized quasiparticle number appear quite different from the switching events associated with charge noise. The discrete jumps have a consistent trend with a positive slope in the magnetic field-gate voltage plane, indicating that they are not an effect due to random charge noise fluctuations. 

Additionally, note that both the isolated discrete jumps in phase and the transition to the low and high field regions occur with approximately the same slope in the $B-V_g$ plane of $\approx 0.4$ V/T. This supports the idea that the discrete phase jumps and the low/high field shifts in behavior are both the result of changing quasiparticle number. The fact that the slope is positive can be understood from the fact that increasing magnetic field is expected to remove quasiparticles (or add quasiholes), while increasing gate voltage should favor adding quasiparticles (or removing quasiholes). It is also noticeable that the lines of constant phase in the high and low field regions are not perfectly flat, but have a slight positive slope. While zero dependence on magnetic field would be expected if all device parameters were independent of magnetic field, the slight positive slope may suggest that increasing $B$ exerts a small negative effective gate voltage due to the increase in cyclotron energy; similar behavior has been observed in quantum dots in the integer quantum Hall regime \cite{Ensslin2020, Ensslin2020-2}.

\section{Supplemental Section 2: Fourier transforms at $\nu = 1/3$}

\begin{figure}[h]
\def\ffile{Figs/v=1_3_FFTs.pdf}
\centering
\includegraphics[width=1\linewidth]{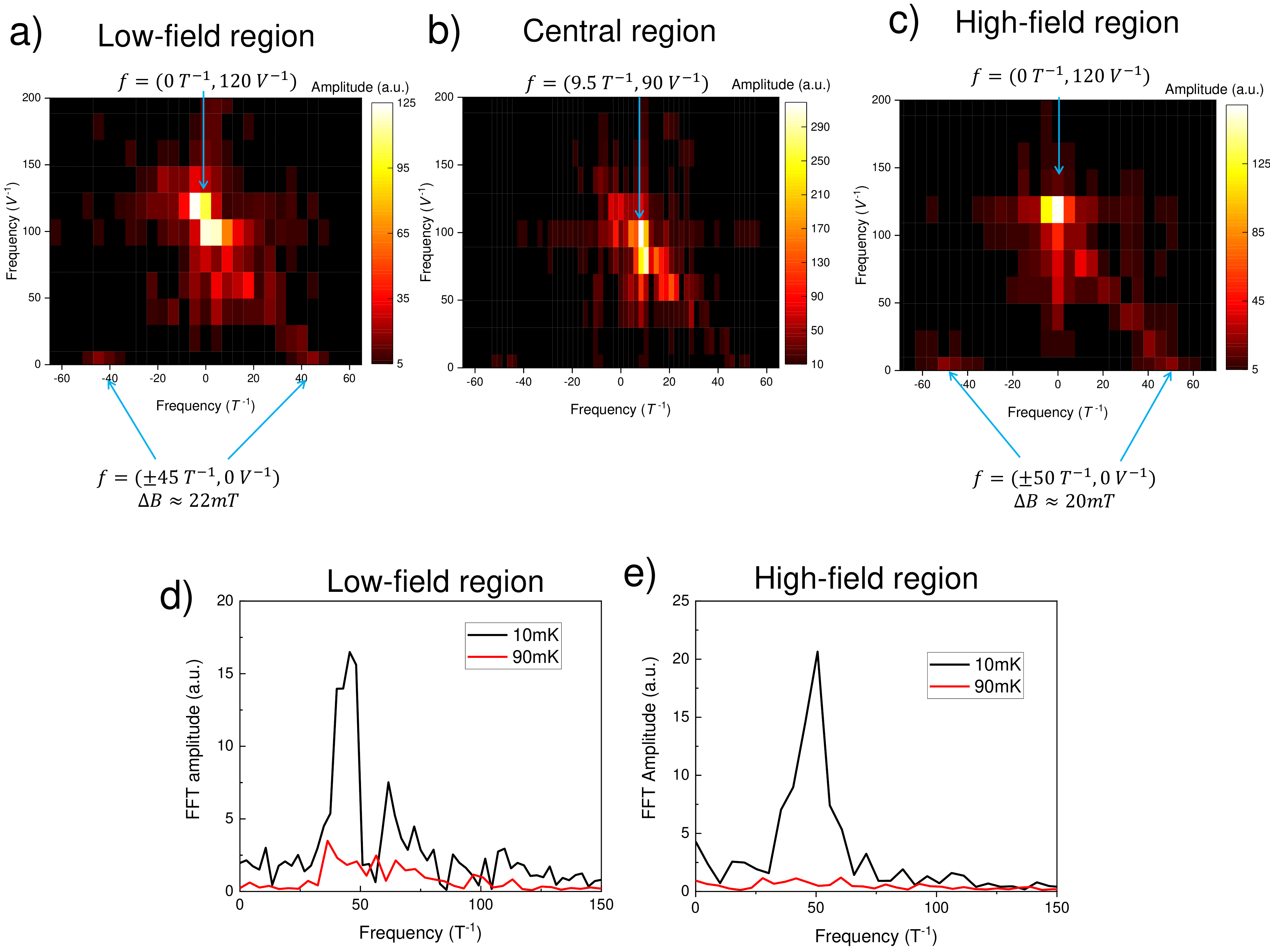}
\caption{\label{FFTs} 2D Fourier transforms of the data in the low-field region from Fig. 1c of the main text in a) the low field region, b) the central region, and c) the high field region. d) Line cuts of the FFT amplitude versus magnetic field frequency at zero gate-voltage frequency, at 10 mK (black) and 90 mK (red) in the low-field region. e) Line cuts at zero gate-voltage frequency in the high-field region. }
\end{figure}

2D Fourier transforms from the data in Fig. 1c from the main text are shown in Fig. \ref{FFTs}. The transform in Fig. \ref{FFTs}a corresponds to the low field region (from 6.95 T to 7.17 T), b corresponds to the central region (7.17 T to 7.65 T), and c corresponds to the high field region (7.65 T to 7.85 T). 

The peak frequencies in the high and low field regions occur at close to 0 magnetic field frequency and 120 V$^{-1}$ gate voltage frequency (corresponding to a period of approximately 8.3 mV). The fact that this peak occurs at zero magnetic field frequency is consistent with expectations for a compressible bulk when $\mu$ is outside of the energy gap \cite{Halperin2021}. In addition to this primary peak, there is a much smaller peak at $\pm 45$ T$^{-1}$ in the low field region and $\pm 50$ T$^{-1}$ in the high field region at 0 gate voltage frequency, corresponding periods of 22 mT and 20 mT. These peaks come from modulations visible in the data (Fig. 1c in the main text), and the periods are close the period of $\approx 20$ mT measured for Aharonov-Bohm interference at $\nu = 1$ (in the compressible region at $\nu=1$ where bulk-edge interaction does not modify the period), indicating that this period corresponds to $\Phi_0$. The $\Phi_0$ period modulations are an expected signature of anyonic braiding statistics when the bulk is compressible \cite{Rosenow2020, Halperin2021} resulting from period changes in localized quasiparticle number. The fact that this peak occurs at zero gate voltage frequency may be a coincidence due to the fact that the lever arm connecting the side gates to the bulk, $\alpha _{buk}$, is approximately half of the lever arm connecting the gate to the edge, $\alpha _{edge}$. The estimated lever arms for the side gates to the edge is $\alpha_{edge} = 0.074$ mV$^{-1}$, and the lever arm for the edge to the bulk is $\alpha _{bulk} = 0.044$ mV$^{-1}$ based on the $\nu = 1$ interference gate voltage period and the $B = 0$ Coulomb blockade period (here the lever arms represent the number of electrons moved per mV change in gate voltage).

It is noteworthy that these modulations have an oscillation period very close to the $\nu = 1$ Aharonov-Bohm period. This strongly indicates that the effective area of the interferometer does not change significantly between $\nu = 1$ and $\nu = 1/3$.

In principle higher frequency harmonics at multiples of $\Phi_0$ might occur, since the change in phase when the quasiparticle number changes is discrete, which would result in a sawtooth-like conductance pattern rather than sinusoidal oscilations \cite{Halperin2011}. However, peaks at these higher harmonics are not visible, likely due to thermal smearing which makes the quasiparticle transitions not sharp. 

Line cuts of the FFT amplitude vs. magnetic field frequency are shown in Fig. \ref{FFTs}d and e for the low and high field regions. The peaks close to the $\Phi_0$ frequency are clearly visible. In red, Fourier transform line cuts at elevated mixing chamber temperature of 90 mK are shown; at this elevated temperature the $\Phi_0$ peaks not visible, consistent with thermal smearing of the quasiparticle number. This reinforces that the $\Phi_0$ modulations are a higher-order contribution to interference which is quickly suppressed as $T$ increases, while the leading order behavior has no $B$ dependence. Note that the charging energy should set the energy scale which determines the visibility of the $\Phi_0$ modulations in the interference pattern at $\nu = 1/3$. According to Ref. \cite{Rosenow2020} the energy scale for thermal damping of the $\Phi_0$ period is $\frac{(e^*)^2 E_c}{\pi^2}$. Using $E_c = 72$ $\mu$eV based on the Coulomb blockade measurements discussed in the main text yields an energy scale of 9.4 mK. Since our dilution refrigerator has a base temperature of approximately 10 mK, it is reasonable that the $\Phi_0$ modulations would be visible for this device at base temperature, but not for larger devices with significantly smaller charging energies or at elevated temperatures.

In the central region the phase evolves primarily due to the Aharonov-Bohm phase, but the isolated discrete jumps also have a noticeable impact on the Fourier spectrum. The peak in the central region occurs at approximately 9.5 T$^{-1}$ and 90 V$^{-1}$. The magnetic field period is lower than the frequency of 12 T $^{-1}$ that would be predicted based only on the oscillation period of 83 mT in the regions between the phase jumps. The shift in frequency occurs due to the discrete jumps in phase, which increase the spacing between peaks and minima in conductance in the regions where they occur, shifting the Fourier peak to lower frequency. 


\section{Supplemental Section 3: Estimating bulk-edge coupling from magnetic field periods}
Due to its smaller size, this device might be expected to have enhanced bulk edge coupling when compared to devices we have studied previously. Enhanced bulk edge coupling should be reflected in the Aharonov-Bohm periods and in the size of the discrete jumps in phase.

In the absence of bulk edge coupling, the interference phase will be given by Eqn. \ref{theta_AB}:
\begin{equation} \label{theta_AB}
\frac{\theta}{2\pi}=e^*_{in} \frac{AB}{\Phi_0}+N_L \frac{\theta_a}{2\pi}
\end{equation}

$B$ is the magnetic field, $A$ is the interferometer area, $e^*_{in}$ is the quasiparticle charge on the interfering edge state, $\Phi_0$ is the flux quantum, $N_L$ is the number of localized quasiparticles, and $\theta_a$ is the anyonic phase. When there is finite bulk edge coupling the phase is modified as given by Eqn. \ref{theta_BE} \cite{VonKeyserlingk2015}:

\begin{equation} \label{theta_BE}
\frac{\theta}{2\pi}=e^*_{in} \frac{\Bar{A} B}{\Phi _0}- \frac{K_{IL}}{K_I} \frac{e^*_{in}}{\Delta \nu} (e^*_{in}N_L + \nu_{in} \frac{\Bar{A} B}{\Phi _0}-\Bar{q})+N_L \frac{\theta_a}{2\pi}
\end{equation}

This modification comes about because there will be variations $\delta A$ in the area due to the bulk edge coupling. $\phi \equiv \frac{\Bar{A}B}{\Phi_0}$ is the flux through the average area $\Bar{A}$ (note that $\Bar{A}$ does not include modulations $\delta{A}$ induced by bulk edge coupling), $\Delta \nu$ is the difference in filling factor between the interfering edge state and the next outer one, $\nu_{in}$ is the filling factor corresponding to the interfering edge state, and $\Bar{q}$ is the background charge (which may be modified by the gate voltage). This implies that if the number of quasiparticles $N_L$ is kept fixed and the background charge is kept fixed, the device will have an oscillation period given by Eqn. \ref{delta_B}:

\begin{equation} \label{delta_B}
\Delta B = \frac{\Phi_0}{e^*_{in} \Bar{A}}(1-\frac{K_{IL}}{K_I}\frac{\nu_{in}}{\Delta \nu})^{-1}
\end{equation}

On the other hand, as derived in \cite{Halperin2011}, assuming there is no cost for creating localized quasiparticles (or electrons for integer states), localized charges will be created which make the average change in bulk charge with $B$ zero, and as long as $\frac{K_{IL}}{K_I}<0.5$ the interference pattern returns to normal Aharonov-Bohm interference (for integer states) with period $\Delta B = \frac{\Phi_0}{e^*_{in} \Bar{A}}$ (although at low temperature there will be modulations due to the phase jumps when quasiparticles enter, even for integer states where there is no anyonic phase). 

\begin{figure}[h]
\def\ffile{Figs/v=1_periodsV3.pdf}
\centering
\includegraphics[width=1\linewidth]{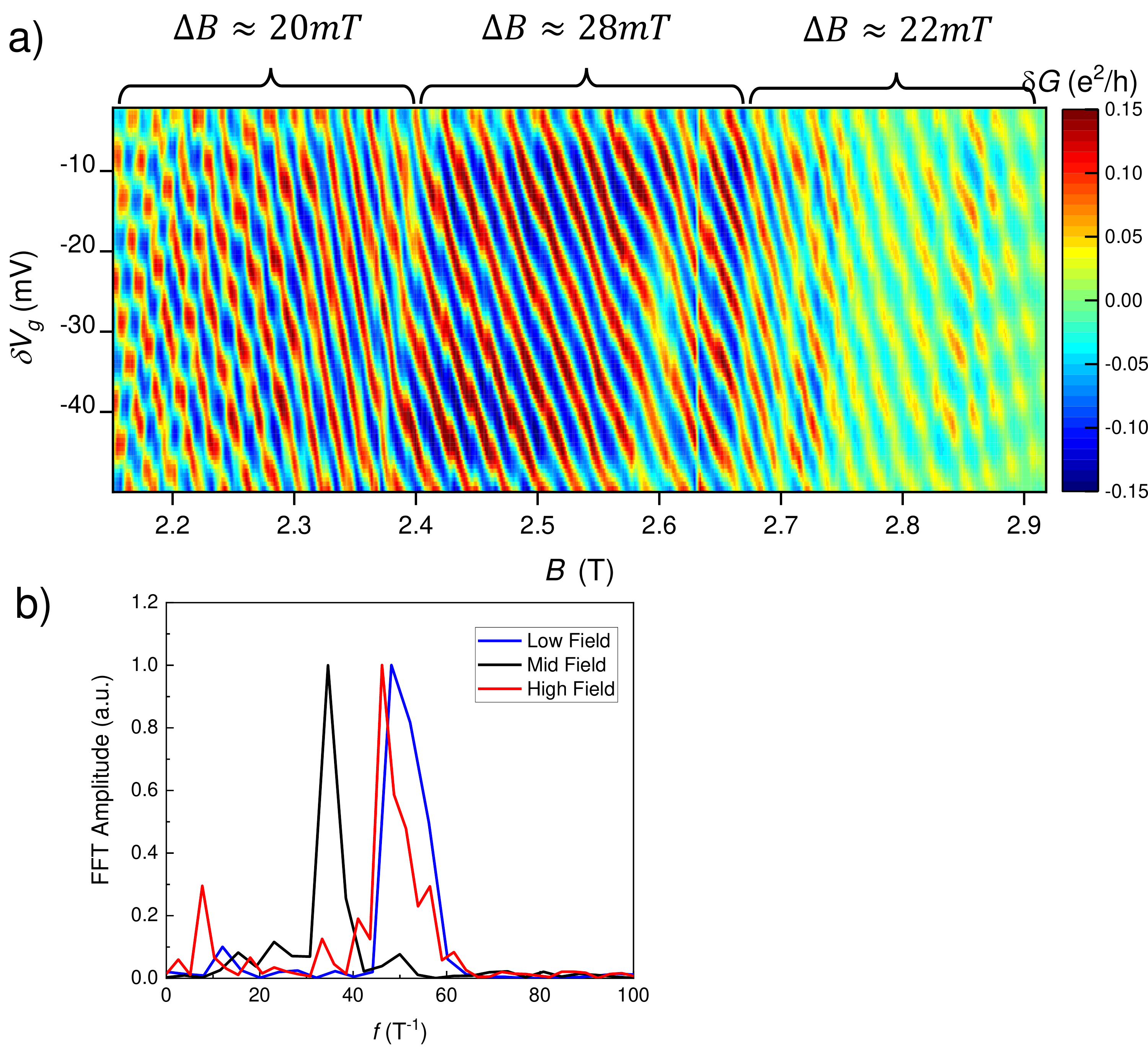}
\caption{\label{v=1} a) Interference across the $\nu=1$. Though less distinct than at $\nu=1/3$, the device appears to show three distinct regions with different magnetic field periods. This suggests that there is a central region where $\mu$ is in the gap and no localized electrons are created, and regions at low and high field where electrons and holes are periodically added to the device. b) Fourier transforms in the low-field, center, and high-field regions. The peak frequencies are used to extract the magnetic fields listed.}
\end{figure}

The theory of \cite{Rosenow2020} predicts that the physics of constant $\nu$ when $\mu$ is in the gap and the bulk is incompressible vs. constant density when the density of states is high and the bulk is compressible should also apply to integer states. Data from our device at $\nu = 1$ appears to show evidence for this: near the center of the plateau there are AB oscillations with negative slope and a period of $\approx 28$ mT, while at lower field and higher field there are oscillations with a somewhat smaller period ($\approx 20$ mT at low field and $\approx 22$ mT at higher field). This is shown in Fig. \ref{v=1}a, and the periods are extracted from Fourier transforms in Fig. \ref{v=1}b. The larger period in the center suggests that the oscillation period is scaled up by the factor $(1-\frac{K_{IL}}{K_I}\frac{\nu_{in}}{\Delta \nu})^{-1}$ as would be predicted by Eqn. \ref{theta_BE} for the case of no change in localized charge, while the smaller periods at high and low field suggest a return to the unscaled period due to high DOS and localized electrons being created as predicted in \cite{Rosenow2020}. This is also supported by the fact that there are periodic modulations visible in the interference pattern in the high and low field regions which are not seen in the center, consistent with shifts in the phase due to periodic changes in localized electron number in those regions (in this case, since $\nu = 1$ is an integer state, these modulations likely occur due to the bulk edge coupling rather than anyonic statistics, and are more prominent in this device due to its smaller size). The ratio of the periods gives $1-\frac{K_{IL}}{K_I}\frac{\nu_{in}}{\Delta \nu}\approx 0.75$ (for the unscaled period the average of the high and low fields regions of 21 mT is used). For $\nu=1$ there is a single edge state so $\nu _{in}=\Delta \nu = 1$, so the estimated $\frac{K_{IL}}{K_I}$ is 0.25. Since this value of $\frac{K_{IL}}{K_I}$ is less than 0.5, the device should be in the Aharonov-Bohm regime \cite{Halperin2011}, which is consistent with the fact that the overall behavior is negatively sloped lines of constant phase.

While this shift in behavior is in some ways similar to what is observed at $\nu = 1/3$, a profound difference is that at $\nu = 1$ in the high and low field regions where quasiholes and quasiparticles are being created, the slope of the lines of constant phase becomes steeper (corresponding to a smaller $B$ period), whereas at $v = 1/3$ in the high and low field regions the slope becomes essentially zero as the lines of constant phase become nearly flat. This points to the important difference between anyons and fermions: the effect of removing a $\nu = 1/3$ anyonic quasiparticle (or adding a quasihole) by increasing $B$ is negative shift in phase because $\theta _a = 2\pi /3$, whereas removing a fermionic electron results in a positive shift in phase because the bulk-edge coupling makes the area increase (giving an increase in the Aharonov-Bohm phase).

A similar analysis can be done at $\nu = 1/3$. In between the phase jumps, the oscillation period is $\approx 83$ mT, which can be set equal to Eqn. \ref{delta_B} with $e^*_{in} = 1/3$. $A$ is know from the high and low field region oscillation periods at $\nu=1$ to be $A=\frac{\Phi_0}{\Delta B}\approx 0.2$ $\mu$m$^2$. Using the oscillation period for the modulations in the high field and low field at $\nu = 1/3$ and assuming these are spaced by $\Phi_0$ yields nearly the same $A$, which is good evidence that these modulations are indeed due to the anyonic phase of quasiparticles introduced with period $\Phi_0$. Also, since $\nu = 1/3$ has a single edge mode, $\Delta \nu = \nu =  1/3$. Using these values for $A$ and $\Delta \nu$ in Eqn. \ref{delta_B} yields $(1-\frac{K_{IL}}{K_I})=0.76$ and $\frac{K_{IL}}{K_I}=0.24$. This is quite close to the $\nu = 1$ value. This is reasonable since both states consist of a single edge mode, so a similar charge redistribution is required to change the area of the interference path, resulting in a similar $K_{IL}$.

\section{Supplemental Section 4: Symmetry of potential drop in finite bias measurements}

For an interferometer the conductance oscillates with the phase, with the conductance varying as $\delta G \propto \cos(\theta)$. When a finite source-drain bias is applied, there are two possibilities to consider: that the potential is applied symmetrically (i.e. the edge state on each side of the device carries an equal amount of the out-of-equilibrium current, with half of the applied bias applied to each edge state) or it is asymmetric (one side carries all or most of the current). For integer states (where $e^*=e$ and $\Delta \nu = 1$) and weak backscattering, when the experiment of measuring differential conductance as a function of gate voltage or magnetic field and source drain bias $V_{SD}$ is performed, the symmetric case results in a checkerboard pattern with $\delta G \propto \cos(\frac{2 \pi AB}{\Phi_0}) \cos(\frac{LeV_{SD}}{2 \hbar v_{edge}})$, whereas for asymmetric potential drop $\delta G \propto \cos(\frac{2 \pi AB}{\Phi_0} - \frac{LeV_{SD}}{\hbar v_{edge}})$. In either case the differential conductance should oscillate as a function of $V_{SD}$. In the symmetric case the product of cosines will result in nodes in the oscillation pattern at $\frac{LeV_{SD}}{2 \hbar v_{edge}}=\pi(n+1/2)$, so that the voltage spacing between nodes $\Delta V_{SD}=\frac{2\pi\hbar v_{edge}}{eL}=\frac{h v_{edge}}{eL}=\frac{\Phi_0 \mathcal{E}}{LB}$. In the fully asymmetric case rather than a checkerboard pattern, pajama stripes with constant sloped lines of constant phase would be expected, but the same expression for $\Delta V_{SD}$ would hold, except that in this case $\Delta V_{SD}$ is the oscillation period as a function of $V_{SD}$. This gives $E_{sp}=\frac{\delta n_I^2}{2} \Delta V_{SD}$.

The intermediate case is also possible and has been reported in graphene interferometers \cite{Deprez2021}. Usually in past experiments in GaAs a checkerboard pattern consistent with symmetric potential drop has been seen \cite{McClure2009, Nakamura2019}. Interestingly, in this device when measuring the differential conductance the symmetry of the applied bias depends on the meausurement circuit and how the potential of the screening wells is set relative to the source and drain contacts. To simplify interpretation we focus on measurements where the potential drop appears to be close to symmetric in the main text.

\begin{figure*}[t]
\def\ffile{Figs/v=1_3_velocity_symmetry.pdf}
\centering
\includegraphics[width=1\linewidth]{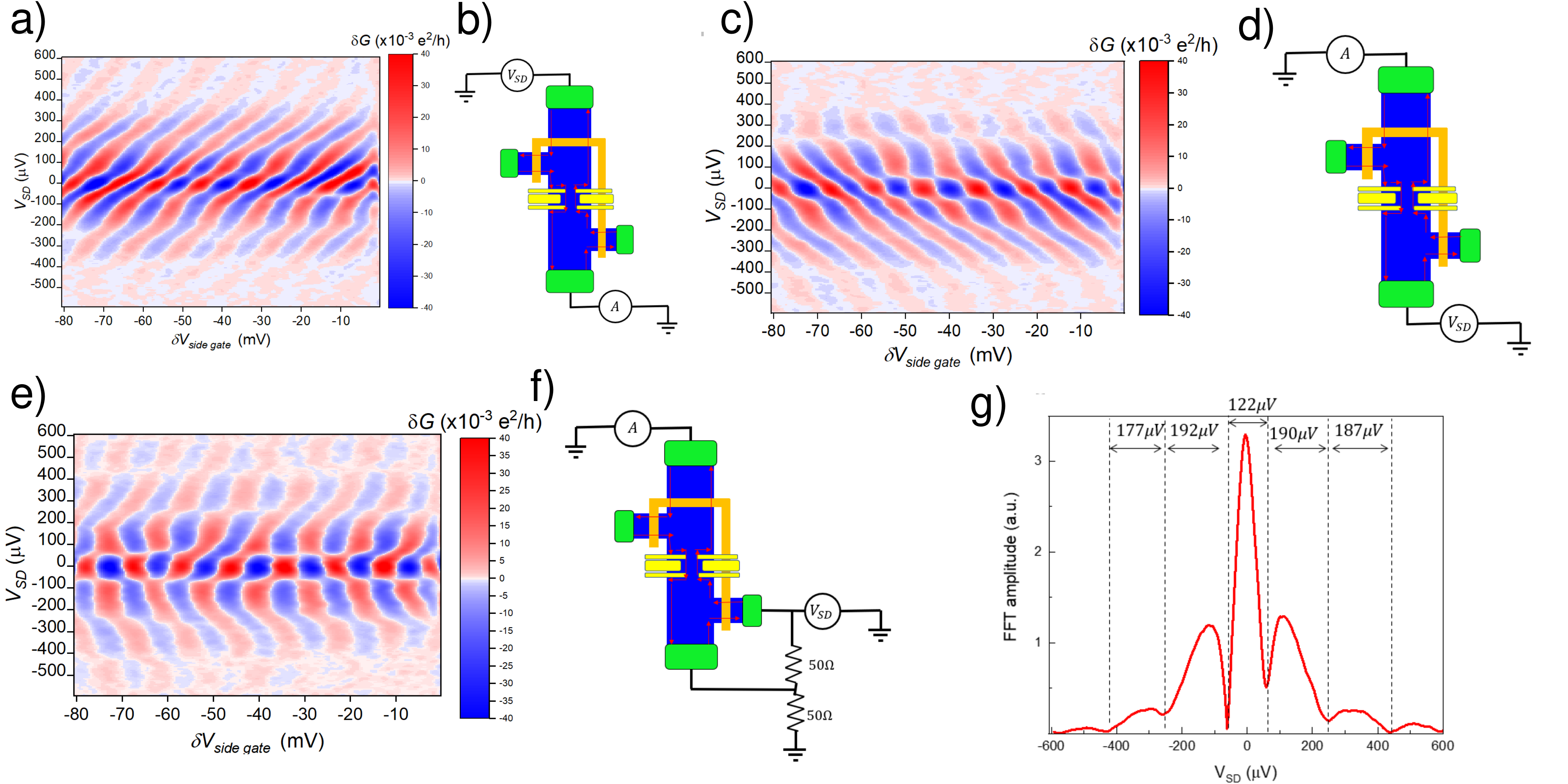}
\caption{\label{v=1_velocity_symmetry} a) Differential conductance measurements at $\nu = 1/3$ $B = 7.4$ T. The overall positive slope suggests that the bulk 2DEG tends to come closer to the potential of the drain contact. b) Schematic showing the cirucit used for the data in a). Green represents the Ohmic contacts, blue represents the mesa and 2DEG, yellow represents the interferometer gates, and orange represents the gate used to isolate the top SW from the Ohmics (there is also an additional gate, not shown, on the reverse side of the chip which isolates the bottom SW from the Ohmics). The bottom Ohmic is not subtended by these gates and thus the SWs equilibrate with this bottom Ohmic. In this circuit setup the bottom contact is used as the drain, with the applied $V_{SD}$ applied to the opposite side of the mesa. c)Differential conductance at $\nu = 1/3$ with the source and drain contacts swapped, with circuit shown in d). e) Differential conductance with non-isolated Ohmic fixed at $V_{SD}/2$, using the circuit shown in f). g) Oscillation amplitude vs. $V_{SD}$. The spacing between nodes (which appear as minima in the plot) are indicated.}
\end{figure*}

In our device, the screening wells are isolated from all of the Ohmic contacts except for one, which prevents any current flowing through the screening wells but ensures that the screening wells are at fixed potential. The single non-isolated ohmic is usually used as a grounding contact; differential conductance data using this configuration is shown in Supp. Fig. \ref{v=1_velocity_symmetry}a, and the circuit is illustrated in Supp. Fig. \ref{v=1_velocity_symmetry}b. It is apparent that there is an overall positive slope to the data, although the data also has some of the character of the checkerboard pattern expected for symmetric potential drop. This suggests that potential in the bulk of the 2DEG inside the interferometer is close to the drain potential rather than being symmetric between the source and drain potential. If the source and drain contacts are switched so that the Ohmic connected to the screening wells is used as the source contact, the slope of the data switches, illustrated in Supp. Fig. \ref{v=1_velocity_symmetry}c and d, indicating that the bulk of the 2DEG is coming close to the potential of the source contact. This indicates that the potential of the screening wells tends to set the electrostatic potential in the bulk of the 2DEG, which is to be expected due to their short setback from the main quantum well.

To achieve the symmetric case, we have implemented a unique biasing scheme in which the non-isolated Ohmic is fixed at 1/2 of the applied $V_{SD}$ (this can be done without directly affecting the current across the device by applied $V_{SD}$ to a downstream Ohmic). The resulting data is shown in Supp. Fig. \ref{v=1_velocity_symmetry}e and the schematic is shown in Supp. Fig. \ref{v=1_velocity_symmetry}f. While the data is fairly symmetric, there is some positive slope behavior when $V_{SD}$ is positive and negative slope when $V_{SD}$ is negative, suggesting that the bulk tends to reach a potential which is somewhat closer to the higher-energy edge state. A similar pattern of different symmetries depending on the source-drain contact configurations occurs at $\nu = 1$, although it is somewhat less pronounced. To simplify interpretation of the data we have analyzed data sets which have this nearly symmetric behavior, which allows the method of extracting velocity from node spacing to be applied as in previous works. The DC current measurements also use the symmetric biasing configuration.

\begin{figure}[t]
\def\ffile{Figs/v=3_outer.pdf}
\centering
\includegraphics[width=1\linewidth]{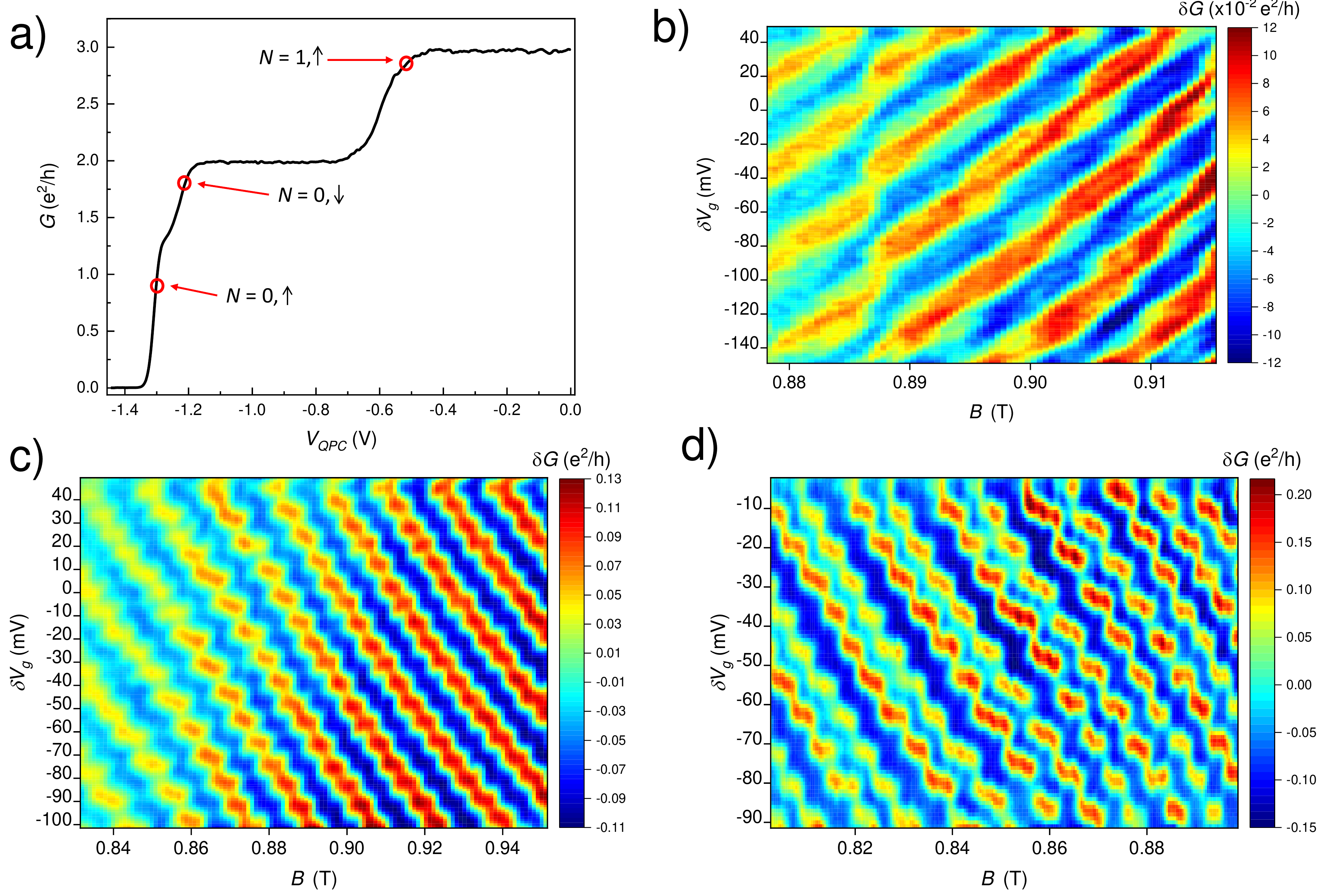}
\caption{\label{v=3_outer} a) QPC sweep at $\nu = 3$. There are clear conductance plateau at $G = 3 \frac{e^2}{h}$ and $G = 2 \frac{e^2}{h}$; there is not a clear $G = 1\times \frac{e^2}{h}$ plateau, likely because the spin gap is too small compared to the cylotron gap, so the two spins of the N = 0 Landau level cannot be completely independently transmitted. However, there is a wiggle in the data that may delineate between primarily backscattering the inner (spin up) N = 0 edge state and primarily backscattering the outer (spin down) N = 0 edge state. Red circles indicate the QPC points where the device is likely primarily partially reflecting a single edge state. b) Repeated from the main text, interference of the innermost edge state corresponding to N = 1, spin up (the corresponding QPC operating point is shown in a). As discussed in the main text, the positive slope indicates Coulomb-dominated behavior, consistent with the value of $\frac{K_{IL}}{K_I} = 0.65 > 0.5$ discussed in the main text. c) Interference of the middle edge state, with QPC operating point also shown in a). The overall negative slope to the data indicates Aharonov-Bohm regime behavior, suggesting that a steeper confining potential towards the outer edge of the sample results in a larger $K_I$. Modulations in the pattern are visible, indicating that effects of bulk-edge coupling are still present. d) Interference of the outermost edge state, with operating point indicated in a). In this case the device exhibits oscillations with approximately half the expected Aharonov-Bohm periods, consistent with previous observations of period-halving for the outermost edge mode when an inner mode is present (and fully reflected). This has been explained by an inter-edge coupling \cite{Rosenow2019}, which is highly plausible because the the two N = 0 edge states are very close together, and also possibly by electron pairing (note the lack of a clear $e^2/h$ plateau in a distinguishing the separate spin edge states supports these two edge states being close together). Clear modulations in the interference pattern are visible suggesting that there may be complicated interplay between bulk-edge coupling and edge-edge coupling.}
\end{figure}

\section{Supplemental Section 5: Finite-bias measurements at elevated temperature}
\begin{figure}[h]
\def\ffile{Figs/v=1_3_high_temp_DC.pdf}
\centering
\includegraphics[width=1\linewidth]{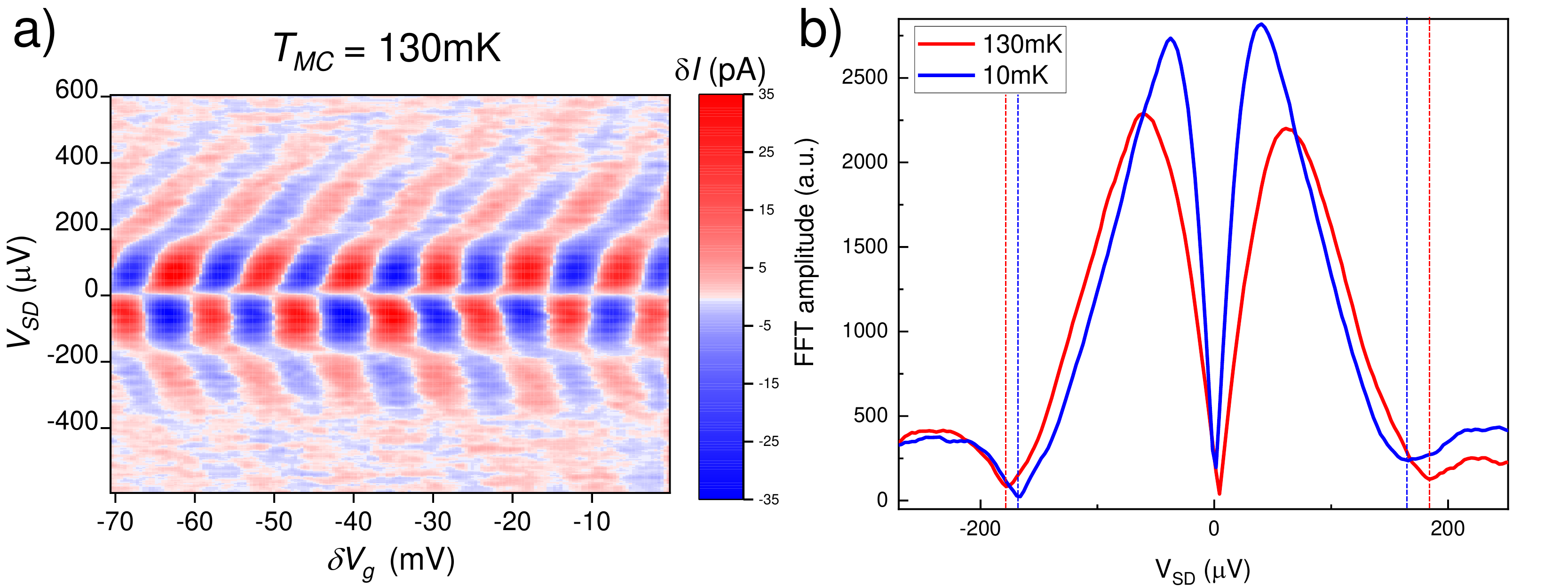}
\caption{\label{v=1_3_high_temp_DC} a) DC current oscillations at $\nu = 1/3$, $B$ = 7.4 T as a function of side gate voltage $\delta V_g$ and $V_{SD}$. b) Oscillation amplitude versus $V_{SD}$ at 10 mK and 130 mK. Dashed lines indicate the positions of the minimum in the oscillations, which correspond to the innermost node. At 130 mK these nodes occur at approximately -178 $\mu$V and +185 $\mu$V, and at 10 mK they occur at -166 $\mu$V and +166 $\mu$V. The fact that these inner nodes move to larger $V_{SD}$ and approach the outer node spacing of $\approx 195$ $ \mu$V at elevated temperatures is consistent with the expected Luttinger-liquid behavior from the model of \cite{Chamon1997}. }
\end{figure}

As discussed in the main text, at low temperature the finite bias current measurements at $\nu = 1/3$ exhibit a non-uniform node spacing in agreement with the predictions of \cite{Chamon1997}. We have also measured oscillations at an elevated mixing chamber temperature of 130 mK, shown in Supp. Fig. \ref{v=1_3_high_temp_DC}. At this elevated temperature the innermost node (corresponding to the minima in the FFT amplitude) moves outward to higher $V_{SD}$ compared to the low-temperature measurement; this comparison is shown in Supp. Fig. \ref{v=1_3_high_temp_DC}b. The inner nodes move from approximately -166 $\mu$V and +166 $\mu$V at 10 mK to -178 $\mu$V and +185 $\mu$V at 130 mK. The outer node spacing is approximately 197 $\mu$V, so at 130 mK the node spacing has become nearly uniform. 

It is also noticeable that at small $V_{SD}$, the oscillation amplitude is much larger at 10 mK mixing chamber temperature than at 130 mK, consistent with thermal dephasing. At elevated $V_{SD}$, however, this difference is less pronounced. This suggests that large $V_{SD}$ causes significant heating of the electrons in the device so that the electron temperature is above the mixing chamber temperature. This might explain why the ratio of inner node spacing is larger than the value of $\frac{1+g}{2} = 2/3$ expected by \cite{Chamon1997} at low temperature, since the applied bias may already cause significant heating and a partial shift towards the high-temperature limit of uniform spacing.

\section{Supplemental Section 6: Calculating phase by Fourier transform}

\begin{figure}[h]
\def\ffile{Figs/v=1_3_phase_analysis_exp_V2.pdf}
\centering
\includegraphics[width=1\linewidth]{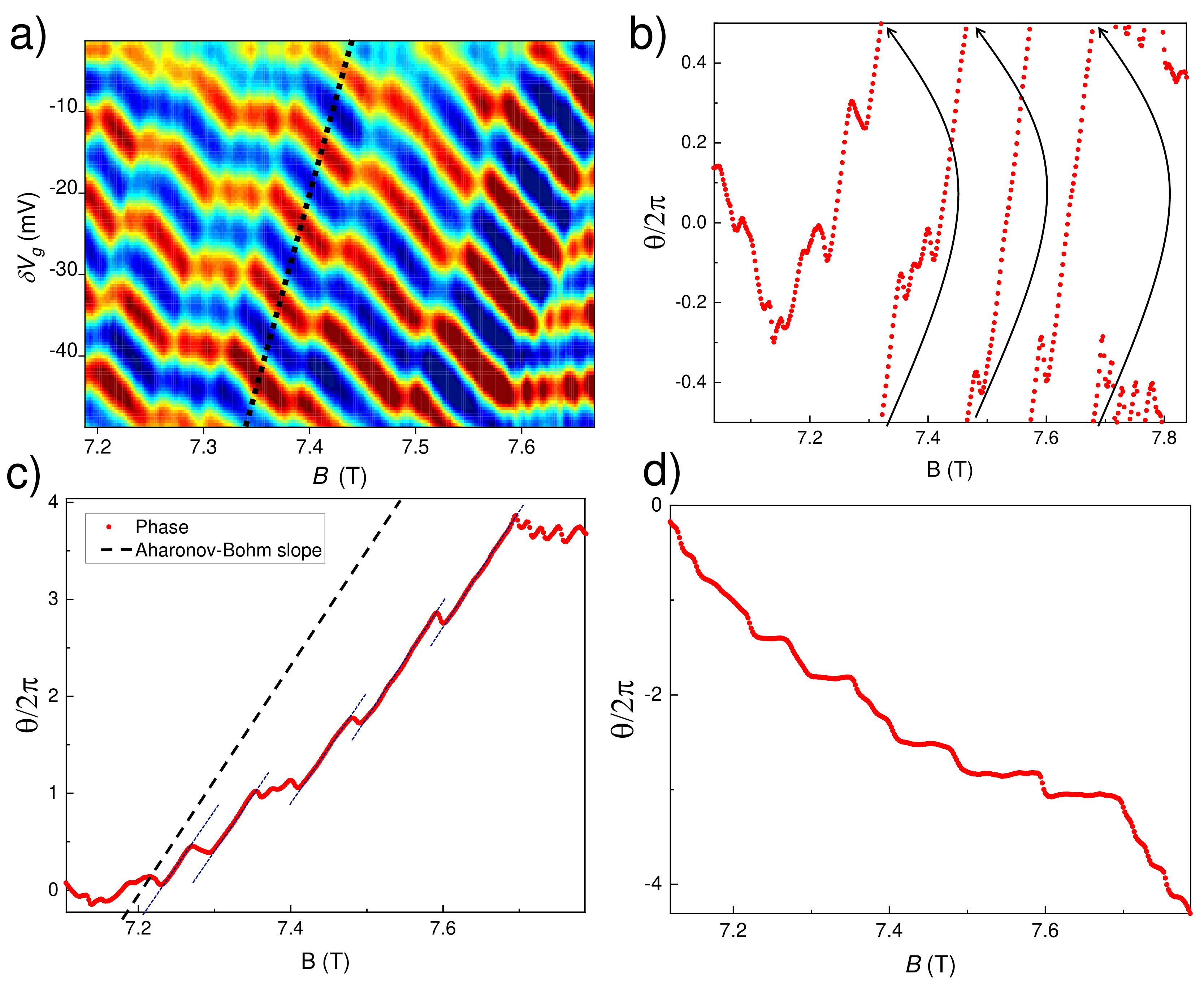}
\caption{\label{phasesExpo} a) Interference data at $\nu = 1/3$; this is a subset of the data shown in the main text. b) Phases extracted from Fourier transforms of the data in a). The phase is evaluated at the peak frequency which corresponds to the Aharonov-Bohm oscillation frequency c)Phase after shifting the phase by $2\pi$ when it crosses from $+\pi$ to $-\pi$ to avoid these discontinuities. The black dashed line indicates the constant Aharonov-Bohm slope, and the blue dashed lines indicate that this slope is consistent in the regions between jumps. d) Phases with the Aharonov-Bohm slope subtracted to isolate the contribution from the discrete phase jumps.}
\end{figure}

In order to accurately calculate the values of the phase jumps that occur at $\nu = 1/3$, we have employed a different method in which the phases are extracted from a Fourier transform of the conductance data. Fourier transforms are used to find the value of the phase $\theta$ at each value of magnetic field. These FFTs are taken along cuts of conductance that are parallel to the discrete jumps in phase so that they do not cross the discrete jumps; this enables the discrete jumps to be made as sharp as possible in plots of $\theta$ versus $B$. The slope of these cuts is 0.4 V/T, as illustrated in Supp. Fig. \ref{phasesExpo}a.

The phases extracted from the FFT, plotted in Supp. Fig. \ref{phasesExpo}b, are defined from $-\pi$ to $\pi$. To remove the discontinuities that occur when crossing this range, the data is shifted up when there is a crossover from $-\pi$ to $+\pi$, as illustrated by the arrows in Supp. Fig. \ref{phasesExpo}b. The resulting phases are plotted in Supp. Fig. \ref{phasesExpo}c.

In the central region where the discrete jumps in phase are mostly well isolated from each other, the phase primarily evolves due to the Aharonov-Bohm effect, which gives a constant phase evolution $\frac{d\theta}{dB} = \frac{2\pi e^* A}{\Phi_0} (1-\frac{K_{IL}}{K_I} \frac{\nu _{in}}{\Delta \nu})$; here the effect of finite bulk-edge interaction is included. This slope can be found by calculating the slope in between the discrete jumps, resulting in an Aharonov-Bohm slope of $\approx 0.012$ T$^{-1}$ (corresponding to a period of 83 mT). This enables extraction of $\frac{K_{IL}}{K_I}$ as discussed in the main text. This slope is shown by the dashed line in the figure, and is consistent across the different regions between the discrete jumps, as expected. In order to isolate the phase contributions from the anyonic statistics found in the discrete jumps, this Aharonov-Bohm slope is subtracted off, with the resulting phase shown in Fig. \ref{phasesExpo}d; this is the same data shown in the main text. Plateaus occur corresponding the regions between the discrete jumps, and the value of the discrete jumps can be computed by the difference in phase from one plateau to another. 



